\documentclass[twocolumn,times]{aastex62}

\usepackage{graphics,graphicx}
\usepackage{psfig}
\usepackage{wrapfig}
\usepackage{picinpar}
\usepackage{graphicx}
    \usepackage{threeparttable}

\def\gax{{$\mathrel{\hbox{\rlap{\hbox{\lower4pt\hbox{$\sim$}}}\hbox{$>$}}}$}}
\def\ltsima{$\; \buildrel < \over \sim \;$}
\def\simlt{\lower.5ex\hbox{\ltsima}}
\def\gtsima{$\; \buildrel > \over \sim \;$}
\def\simgt{\lower.5ex\hbox{\gtsima}}
\newcommand{\msun}{{\rm\,M$_\odot$}}

\begin{document}

\title{
Transient radio emission from low-redshift galaxies at $z<0.3$ revealed by VLASS and FIRST surveys 
}
\correspondingauthor{Xinwen~Shu \& Luming~Sun} 
\email{xwshu@ahnu.edu.cn, sunluming@ahnu.edu.cn}

\author{Faobao~Zhang}
\affil{Department of Physics, Anhui Normal University, Wuhu, Anhui 241002, China}

\author[0000-0002-7020-4290]{Xinwen~Shu}
\affil{Department of Physics, Anhui Normal University, Wuhu, Anhui 241002, China}

\author{Luming~Sun}
\affil{Department of Physics, Anhui Normal University, Wuhu, Anhui 241002, China}

\author{Lei~Yang}
\affil{Department of Physics, Anhui Normal University, Wuhu, Anhui 241002, China}

\author{Ning Jiang}
\affil{Department of Astronomy, University of Science and Technology of China, Hefei, Anhui 230026, China}

\author{Liming~Dou}
 \affiliation{Department of Astronomy, Guangzhou University, Guangzhou 510006, China}

\author{Jianguo~Wang}
\affiliation{Yunnan Observatories, Chinese Academy of Sciences, Kunming 650011, China}

\author{Tinggui~Wang}
\affil{Department of Astronomy, University of Science and Technology of China, Hefei, Anhui 230026, China}

\begin{abstract}

We present the discovery of a sample of 18 low-redshift ($z<0.3$) galaxies with transient {nuclear} radio emission. 
These galaxies are not {or weakly} detected 
in the Faint Images of the Radio Sky at Twenty cm survey performed on {1993-2009}, but have brightened significantly 
in the radio flux {(by a factor of $\simgt$5}) in the epoch I (2017-2019) observations of Very Large Array Sky Survey (VLASS). 
{All the 18 galaxies} have been detected in the epoch II VLASS observations in 2020-2021, 
for which the radio flux {is found to evolve slowly (by a factor of $\simlt$40\%}) over a period of about three years.  
{15 galaxies have been observed in the Rapid ASKAP Continuum Survey, 
and a flat or inverted spectral slope between 888 MHz and 3 GHz is found.}
Based on the Sloan Digital Sky Survey spectra taken before the radio brightening, 
14 out of 18 can be classified to be {LINERs or normal galaxies with weak or no nuclear activity. }
Most galaxies are red and massive, with {more than} half having central black hole masses above $10^{8}$\msun. 
We find that only one galaxy in our sample displays optical flare lasting for at least two months, 
and a long decay in the infrared light curve that can be explained as the dust-heated echo emission of central optical flare, such 
as {a stellar} tidal disruption event. 
We discuss several possibilities for the transient radio emission and conclude that 
it is likely associated with a new-born radio jet triggered by short sporadic fueling of supermassive black hole. 
Such a scenario can be tested with further multi-frequency radio observations 
{of these sources through measuring their radio flux variability and spectral evolution.} 

\end{abstract}


\section{Introduction} \label{sec:Intro}

Relativistic jets have been observed from stellar-mass black holes ($M_{\rm BH}$$\sim$10\msun) in Galactic 
X-ray binaries (XRBs), to supermassive black holes (SMBHs, $M_{\rm BH}$$\sim$$10^6-10^9$\msun) 
in the centers of most galaxies \citep{Blandford2019}. It has been established that the form of jet in an XRB is determined 
by the accretion state. 
During the low/hard state a steady, compact radio jet is produced, which is however significantly 
quenched in the high/soft state where disk thermal emission is dominant \citep{Fender2004}. 
As sources transition from low to high accretion state, 
isolated radio flares associated with the launch of relativistic ejecta have also been observed \citep{Hjellming1995, Bright2020}. 
However, such a ubiquitous feature of jet {production and quenching} regulated by accretion states 
has rarely been seen in active galactic nuclei (AGN). 
{This is because several criteria must be met by the source at once, and such AGNs are not common. 
First, the X-ray variability should be large enough in order to identify the accretion model transition 
\citep[e.g.,][]{Liu2020, Ricci2021}. Second, there must be a well-measured radio emission at different 
accretion states so that its evolution can be used to constrain the jet formation and quenching. 
}
Although ``changing-look'' AGN are ideal targets for such studies owing to large changes 
in accretion rates, no jet-related transient radio emission has been convincingly detected so far \citep{Gezari2017, Yang2021}. 
Furthermore, it remains mysterious why only a small fraction of AGNs ($\sim$10\%) 
{are radio-loud associated with relativistic jets \citep[e.g.,][]{Kellermann2016, Blandford2019}. 
While \citet{Moravec2022} recently show that different radio AGN populations may follow an 
evolutionary track in the X-ray hardness-luminosity diagram similar to XRBs, it is difficult to 
directly correlate them with specific accretion state transitions, as their X-ray spectra are still dominated 
by coronal emission. 
}

\begin{table*}
\tablewidth{0pt}
\centering
\caption{Observation log and radio flux measurements}
\setlength{\tabcolsep}{1.5mm}
\begin{tabular}{cccccccclcl}\hline\hline
Name & R.A. & Decl. & $z$ & Obs. Date & Peak Flux  & Int. Flux & log($\nu$$L_{\rm 3GHz}$) &  Obs. Date & Int. Flux \\
 &  &  &  & (VLASS) & (mJy/beam) & (mJy) & (erg $s^{-1}$) &   (FIRST) & (mJy) \\\hline
J0040+0823 & 10.1845 & 8.3978 & 0.214 & 2017-11-27 & $5.30 \pm 0.07$ & $5.20\pm 0.13$ & 40.33 &  2009-04 & $0.80 \pm 0.12$ \\
 &  &  &  & 2020-09-19 & $6.61 \pm 0.08$ & $6.55\pm 0.15$ & 40.43 &  &  & \\
J0154-0111 & 28.5486 & -1.1971 & 0.046 & 2017-09-27 & $5.70 \pm 0.13$ & $5.88\pm 0.23$ & 38.95 &  1995-11 & $0.58 \pm 0.14$ \\
 &  &  &  & 2020-08-15 & $6.13 \pm 0.12$ & $6.42\pm 0.21$ & 38.99 &  &  & \\
J0800+2928 & 120.0671 & 29.4714 & 0.045 & 2019-04-13 & $9.99 \pm 0.07$ & $10.09\pm 0.11$ & 39.17 &  1993-05 & $<$ 0.44  \\
 &  &  &  & 2021-11-29 & $14.09 \pm 0.16$ & $14.44 \pm 0.28$ & 39.33 &  &  & \\
J0950+5128 & 147.6532 & 51.4772 & 0.211 & 2019-04-18 & $8.94 \pm 0.08$ & $8.24 \pm 0.13$ & 40.51  &  1997-04 & $<$ 0.41 \\
 &  &  &  & 2021-11-22 & $11.16 \pm 0.06$ & $10.82 \pm 0.12$ & 40.63 &  &  & \\
J0951+3703 & 147.9234 & 37.0596 & 0.236 & 2019-04-26 & $9.26 \pm 0.11$ & $9.46\pm 0.20$ & 40.69 & 1994-07 & $<$ 0.42 \\
 &  &  &  & 2021-12-06 & $10.69 \pm 0.15$ & $10.44 \pm 0.25$ & 40.73  &  &  & \\
J1029+0436 & 157.4603 & 4.6161 & 0.085 & 2017-11-26 & $5.82 \pm 0.07$ & $5.66\pm 0.12$ & 39.49 &   2000-02 & $<$ 0.44 \\
 &  &  &  & 2020-08-08 & $7.13 \pm 0.20$ & $7.13 \pm 0.20$ & 39.65 &  &  & \\
J1129+3900 & 172.4168 & 39.0129 & 0.287 & 2019-05-04 & $11.39 \pm 0.14$ & $11.38 \pm 0.24$ & 40.94 &  1994-08 & $0.46 \pm 0.14$ \\
 &  &  &  & 2021-12-06 & $10.71 \pm 0.11$ & $10.75 \pm 0.19$ & 40.92 &  &  & \\
J1217+2750 & 184.3733 & 27.8413 & 0.184 & 2017-11-24 & $7.33 \pm 0.10$ & $7.79\pm 0.18$ & 40.36 & 1995-11 & $0.31 \pm 0.14$ \\
 &  &  &  & 2020-09-06 & $5.02 \pm 0.43$ & $5.84\pm 0.84$ & 40.23 &  &  & \\
J1301+2127 & 195.3815 & 21.4636 & 0.087 & 2017-09-25 & $7.68 \pm 0.21$ & $8.09\pm 0.39$ & 39.67 &  1998-10 & $<$ 0.43 \\
 &  &  &  & 2020-07-16 & $7.29 \pm 0.09$ & $7.28\pm 0.16$ & 39.62 &  &  & \\
J1337+3857 & 204.4213 & 38.9587 & 0.243 & 2017-10-14 & $3.36 \pm 0.14$ & $4.21\pm 0.30$ & 40.36 &  1994-08 & $<$ 0.68 \\
 &  &  &  & 2020-09-17 & $2.58 \pm 0.07$ & $2.93\pm 0.13$ & 40.21 &  &  & \\
J1407+1247 & 211.9427 & 12.7889 & 0.126 & 2019-04-17 & $6.89 \pm 0.13$ & $6.39\pm 0.21$ & 39.91 &   1999-12 & $0.97 \pm 0.14$ \\
 &  &  &  & 2021-10-01 & $6.33 \pm 0.31$ & $6.32 \pm 0.54$ & 39.91 &  &  & \\
J1409+5420 & 212.4675 & 54.3485 & 0.174 & 2017-12-01 & $5.58 \pm 0.08$ & $5.74\pm 0.14$ & 40.17 &  1997-05 & $<$ 0.42 \\
 &  &  &  & 2020-09-01 & $5.74 \pm 0.17$ & $5.59\pm 0.28$ & 40.16 &  &  & \\
J1437-0033 & 219.3984 & -0.5567 & 0.180 & 2019-05-01 & $6.16 \pm 0.12$ & $6.38\pm 0.21$ & 40.25 &  1998-08 & $<$ 0.43 \\
 &  &  &  & 2021-10-24 & $6.49 \pm 0.24$ & $6.79 \pm 0.43$ & 40.28 &  &  &  \\
J1558+1412 & 239.6987 & 14.2037 & 0.034 & 2019-04-11 & $28.67 \pm 0.64$ & $29.20\pm 1.10$ & 39.38 &  2000-01 & $<$ 0.40 \\
 &  &  &  & 2021-11-15 & $ 26.72 \pm 0.22$ & $27.27 \pm 0.39$ & 39.35 &  &  & \\
J1610+0606 & 242.5899 & 6.1162 & 0.156 & 2019-03-18 & $5.67 \pm 0.15$ & $5.53\pm 0.25$ & 40.05 & 2000-02 & $<$ 0.45 \\
 &  &  &  & 2021-10-17 & $ 4.57 \pm 0.10$ & $4.47 \pm 0.18$ & 39.96 &  &  & \\
J1642+3346 & 250.6908 & 33.7777 & 0.136 & 2017-10-06 & $6.73 \pm 0.18$ & $6.86\pm 0.31$ & 40.02 &  1994-06 & $0.79 \pm 0.14 $ \\
 &  &  &  & 2021-09-16 & $4.74 \pm 0.11$ & $ 4.43 \pm 0.18$ & 39.83 &  &  & \\
J1646+4227 & 251.5293 & 42.4604 & 0.050 & 2019-05-04 & $9.63 \pm 0.14$ & $9.61\pm 0.25$ & 39.22 &  1995-10 & $<$ 0.40 \\
 &  &  &  & 2021-11-14 & $9.12 \pm 0.06$ & $8.84 \pm 0.10$ & 39.18 &  &  & \\
J2301+0544 & 345.4323 & 5.7387 & 0.140 & 2017-10-23 & $7.09 \pm 0.15$ & $7.53\pm 0.28$ & 40.08 &  1995-10 & $0.47 \pm 0.12 $ \\
 &  &  &  & 2020-08-12 & $7.42 \pm 0.09$ & $7.35\pm 0.16$ & 40.07 &  &  & \\\hline
\end{tabular}
 
\end{table*}

In the past two decade, time-domain surveys have led to the discovery of a population of energetic nuclear transients in otherwise 
quiescent galaxies. 
They are mostly due to the stellar tidal disruption events (TDEs) by SMBHs. 
As the stellar debris gets accreted effectively, a fraction
of accretion power could be converted into outflow, and under certain conditions producing 
a relativistic jet, which can be detected at radio wavelengths \citep[e.g.,][]{Velzen2011, Zauderer2011}. 
Sw J1644+57 is the first and prototype TDE displaying a relativistic jet \citep{Burrows2011}, 
from which a luminous and variable radio emission is readily detected after the stellar disruption \citep{Berger2012}, 
presenting a natural SMBH analogy with state changes of XRBs. 
However, despite extensive searches, radio observations of other TDEs 
have not yet produced conclusive detections of powerful jet as that in Sw J1644+57 \citep{Bower2013, Velzen2013, Velzen2016, Dai2020, Alexander2020}. 
It is suggested that the non-relativistic outflows may be more ubiquitous than jets in TDEs \citep{Alexander2016, Alexander2017, Anderson2020, Mohan2021}, 
but due to insufficient sensitivity of radio observations, most of them cannot be detected \citep{Velzen2016}. 
The current lack of detections can also be explained by the delayed onset
of the radio emission \citep{Horesh2021a, Horesh2021b}. 

Recent radio surveys such as Caltech-NRAO Stripe 82 Survey \citep[CNSS,][]{Mooley2016} and Very Large Array Sky Survey \citep[VLASS,][]{Lacy2020} have opened a new window 
to select TDE candidates as well as other types of nuclear radio transients.  
Using the CNSS data, \citet{Anderson2020} presented the discovery of the TDE candidate   
CNSS J0019+00, through the detection of its transient radio emission, possibly originating from the 
interaction of the non-relativistic outflow with the surrounding medium. 
\citet{Ravi2021} reported the discovery of the fading radio transient FIRST J1533+2727, 
which can be described as the long-lasting radio afterglow of a TDE. 
On the other hand, brightened radio emission has been found in a sample of galaxies and quasars, 
by crossmatching the data taken from the Faint Images of the Radio Sky at Twenty cm survey \citep[FIRST,][] {Becker1995,Helfand2015} with either CNSS \citep{Kunert2020, Wolowska2021} or VLASS \citep{Nyland2020}. 
Since the hosts for most of these radio transients show powerful AGN activities in the preflare optical spectra, their transient 
radio emission has been explained as the transition from radio-quiet to radio-loud state, possibly associated 
with a new-born jet. Follow-up multi-frequency VLA observations suggested that the radio emission is compact with 
a typical source size less than 0\farcs1 ($<$1 kpc), and characterized by a curved radio {spectral energy distribution (SED)} peaking at $\sim$5-10 GHz \citep{Nyland2020, Wolowska2021}, 
making these sources to be consistent with the population of young radio AGNs \citep[][and references therein]{Odea2021}. 

In this paper, we report the identification of a sample of nearby galaxies ( $z<0.3$) with nuclear radio {transients} using the VLASS data.   
Section 2 describes observations and data, and Section 3 presents the selection of radio transients. 
The host galaxies and variability properties at optical and {mid-infrared (MIR)} bands are analyzed in Section 4. 
{In Section 5, we present the discussion on the possible origins of transient radio emission. 
Section 6 is the summary of our findings. }

\section{Observation and Data} \label{sec:Observation and Data}

The FIRST survey was conducted at a frequency of 1.4 GHz, with an angular resolution of 5$\arcsec$ and imaging rms  down to 130 $\mu$Jy/beam in the deepest footprint. It covers 10,575 $\rm deg^2$, approximately 25$\%$ of the total sky area. The first epoch FIRST survey was carried out during the period of 1993-2004, and extended to cover larger sky region with observations performed between 2009 and 2011. Given the coordinates of sources of interest, the image cutouts can be extracted from the FIRST data archive\footnote{http://sundog.stsci.edu/}. 

VLASS is a S-band (2–4 GHz) multi-epoch legacy survey aiming at to detect various types of extragalactic radio transients, such as TDEs, 
accretion state changing AGNs, and afterglows of off-axis $\gamma$-ray bursts and core-collapse supernova \citep{Lacy2020}. The angular resolution of VLASS is about 2$\farcs$5, and it is currently surveying the entire northern sky with Dec$>$ -$40^\circ$ (33,885 $\rm deg^2$). VLASS observations are designed to map the same survey area three times,  
{each separated by approximately a period of 32 months}. Each VLASS epoch achieves an 1$\sigma$ sensitivity of $\sim$120 $\mu$Jy/beam, which is comparable to the depth of FIRST. The combination of images taken from three-epoch VLASS observations will achieve an 1$\sigma$ sensitivity of  $\sim$70 $\mu$Jy/beam, making it the deepest wide-field survey at $\sim$3 GHz after the final implementation of survey plan.   
The VLASS program has began in 2017 and recently completed its first epoch observations in 2019 (epoch I).
{
The preliminary ``QuickLook" images have been publicly released on the NRAO website\footnote{https://archive-new.nrao.edu/vlass/quicklook/}, 
in order to help the scientific community to timely access the VLASS data. 
\citet{Gordon2021} have produced the source catalog from the epoch I VLASS ``QuickLook" images using the 
Python Blob Detector and Source Finder \citep[{\tt PyBDSF},][]{Mohan2015}, consisting of $1.7\times10^{6}$ 
individual radio sources with a peak flux of $\simgt1$ mJy/beam. 
The catalog and an associated User Guide are available at the CIRADA website\footnote{https://cirada.ca/catalogues.}.

\section{Selection of radio transients} \label{sec:selection}

The main goal of present study is to select galaxies that have recently brightened in radio, i.e.,  bright 
radio transient sources (RTS) in VLASS but without counterparts in the previous FIRST catalog ($S_{\rm 1.4 GHz}<1$mJy). 
We first cross-matched the FIRST and epoch I VLASS catalog \citep{Gordon2021}  to search for VLASS sources that 
are not detected in FIRST, within a matching radius of $r=2\farcs$5. 
We restricted the sample to include only sources that have a 3 GHz flux of $\simgt$5 mJy in the VLASS catalog. 
This ensures that the variability amplitude is at least by a factor of 5 in comparison with the FIRST upper limits, 
and also helps to eliminate spurious sources. 
\citet{Gordon2021}  suggested that the detected radio components with a {peak} flux above 5 mJy/beam tend to be more 
reliable in the VLASS ``QuickLook" images (Quality\_flag == 0 in the catalog). 
Then we cross-matched the sample with the SDSS spectroscopic catalog, and required all objects to be nearby 
at $z<0.3$ in order to give efficient classifications of sources using the Baldwin, Philips and Terlevich (BPT) diagram 
if emission lines are detected. 
VLASS sources have been inspected visually to ensure that they are not artifacts such as
extended emission from brighter foreground galaxies, image artifacts, and side lobes due to imperfect 
 complex gain solutions. This leaves a total of 20 VLASS sources. 
 
 \begin{figure}[htbp!]
 \begin{center}
 \includegraphics [width=0.5\textwidth]{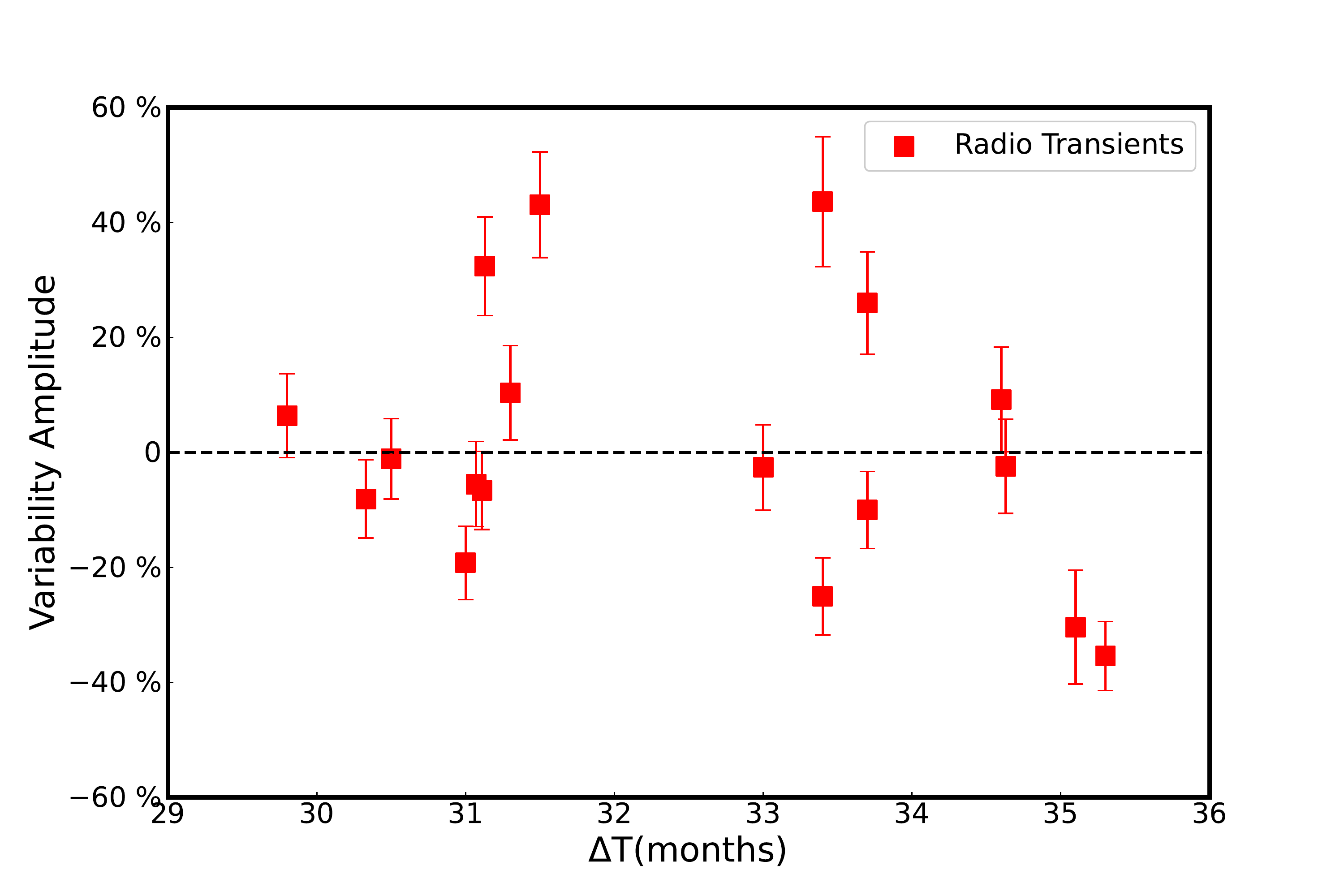}
 \hspace{0.5cm}
 \end{center}
 \vspace{-0.7cm}
 \caption{The percent change in radio flux over the period of two-epoch VLASS observations, 
 which is defined as $(F_{\rm epoch~II}-F_{\rm epoch~I})/F_{\rm epoch~I}$. 
The flux errors include 5\% uncertainty in the flux densities measured from CASA, to account for the flux calibration uncertainty of VLASS data \citep{Lacy2020}.
 } 
 \label{Fig:fluxvar}%
 \vspace{0.2cm}
 \end{figure}

{Since the present work aims to investigate the transient radio emission that is likely associated with SMBH accretion, 
we need to exclude known supernova explosions. 
We cross-correlated with the open supernovae (SNe) catalogue compiled by \citet{Guillochon2017}\footnote{The catalog is 
constructed by the data presented in the SNe literature as well as other web-based catalogs, and available on the website https://sne.space, 
and updated to include all supernovae (and candidates) reported up to Jan 2022. } 
which includes 36,000+ supernovae and related candidates. 
The cross-match results in two known supernovae, J122247.59+053624.3 \citep[SN 2012ab,][]{Bilinski2018} and J131341.47+471756.7 \citep[PTF11qcj,][]{Palliyaguru2019}.  
Note that the VLASS epoch I data for the latter SNe J131341.47+471756.7 has been reported in \citet{Stroh2021}. 
After removing the two known supernovae, our final sample consists of {a total of} 18 galaxies with 
transient radio emission, for which detailed analysis will be presented in this paper. 
}


\begin{figure*}[htbp]
\centering
\includegraphics[scale=0.28]{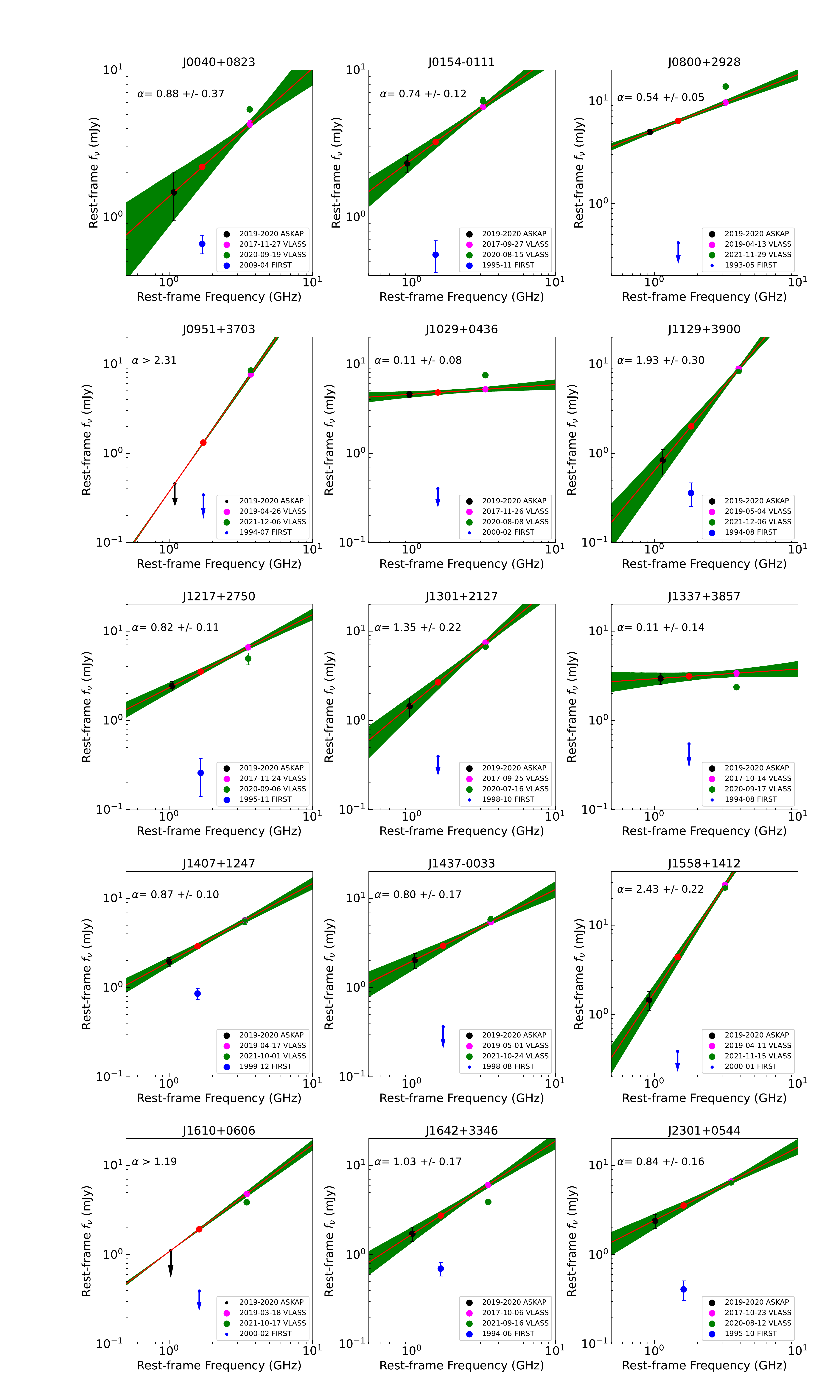}
\caption{
Radio SED between 888 MHz and 3 GHz. 
For sources that are not detected in FIRST or {RACS}, 3$\sigma$ upper limits on flux are shown. 
The error bars for the radio spectral index ($\alpha_{\rm 0.89-3 GHz}$) 
are estimated using Monte Carlo simulations (green shaded regions), assuming that the error on each flux follows Gaussian distribution. 
Filled red circle in each panel represents the extrapolated 1.4 GHz flux density based on the best-fit radio slope. 
 }
 \label{Fig:fitslope}%
\end{figure*}

We used {{\tt IMFIT} task in the} CASA software \citep[version 5.3.0,][]{McMullin2007} to measure the integrated and peak flux for each galaxy. 
The ratio of integrated flux to peak flux is in the range 0.93--1.25, with a median value of 1.02, suggesting that most, if
not all, radio emission is unresolved and compact at the resolution of 2$\farcs$5. We checked that the 
positional offsets of radio sources relative to the optical centers {obtained from the SDSS photometry} 
are in the range 0$\farcs$08--0$\farcs$41, with a median offset of 0$\farcs$2, 
{Note that the positional offsets are comparable to the astrometric accuracy of VLASS \citep[$<$0$\farcs$4,][]{Lacy2020} 
and that of SDSS \citep[$<$0$\farcs$15,][]{York2000}, implying that the radio emission originates from a region close to the galactic center.  
Although the 18 sources are not listed in the FIRST catalog, in order to secure their radio faintness during the FRIST observations, 
we used the CASA software to measure the flux of each source in the FIRST images. 
Given the positions of radio components detected by VLASS, we found that 7 sources are detectable at $\sim3-7\sigma$ with a flux 
in the range 0.46--0.97 mJy\footnote{Note that the 
CLEAN bias correction \citep{White1997} was not considered in calculating the source detection significance. }, while only upper limits can be obtained for the remaining 11 sources, i.e., the detection significance is $<3\sigma$.  
This confirms that the selected VLASS sources are indeed faint ($<$1 mJy) during the FIRST observations. 
All the 18 sources have also been detected in the VLASS epoch II observations.  
Considering the flux errors \citep[including the 5\% uncertainty in the flux calibration,][]{Lacy2020}, the flux densities between 
the two VLASS observations are consistent with each other for 10 sources (Figure \ref{Fig:fluxvar}). 
The radio flux has been found to increase moderately (within a factor of 50\%) in four galaxies, 
while it declines in four galaxies with the variability amplitude in the range $20\%-50\%$. 
The source name, radio coordinate, redshift, VLASS observation date(s) and radio flux measurements, FIRST observation date 
and flux measurements are reported in Table 1. 
The FIRST, VLASS epoch I and epoch II images are shown in Appendix A. 
{Note that the amount of radio flux variations between the two VLASS observations is consistent 
with other samples of radio transient sources \citep{Nyland2020, Wolowska2021}, as shown in Appendix B. }
}

\begin{table}
\centering
\setlength{\tabcolsep}{1.2mm}
\caption{Radio flux and spectral measurements with {RACS} observations}
\begin{tabular}{ccccc}\hline\hline
Name & Peak Flux & Int. Flux & Rms &$\alpha_{\rm0.89-3 GHz}$ \\
& (mJy/beam) & (mJy) &  (mJy/beam) \\\hline
J0040+0823 & 1.78 & 1.71 & 0.64 & 0.88 $\pm$ 0.37\\
J0154-0111 & 2.42 & 2.84 & 0.33 & 0.74 $\pm$ 0.12\\
J0800+2928 & 5.23 & 4.99 & 0.22 & 0.54 $\pm$ 0.05\\
J0951+3703 & $<$0.57 &$\sbond$ & 0.19 & $>$2.31\\
J1029+0436 & 4.96 & 6.02 & 0.38 & 0.11 $\pm$ 0.08\\
J1129+3900 & 1.07 & 1.07 & 0.34 & 1.93 $\pm$ 0.30\\
J1217+2750 & 2.88 & 2.93 & 0.35 & 0.82 $\pm$ 0.11\\
J1301+2127 & 1.57 & 1.75 & 0.38 & 1.35 $\pm$ 0.22\\
J1337+3857 & 3.67 & 6.43 & 0.53 & 0.11 $\pm$ 0.14\\
J1407+1247 & 2.20 & 2.17 & 0.25 & 0.87 $\pm$ 0.10\\
J1437-0033 & 2.40 & 3.74 & 0.46 & 0.80 $\pm$ 0.17\\
J1558+1412 & 1.50 & 2.67 & 0.36 & 2.43 $\pm$ 0.22\\
J1610+0606 & $<$1.29 &$\sbond$ & 0.43 & $>$1.19\\
J1642+3346 & 1.95 & 2.78 & 0.36 & 1.03 $\pm$ 0.17\\
J2301+0544 & 2.73 & 3.86 & 0.49 & 0.84 $\pm$ 0.16\\
\hline
\end{tabular}
\end{table}

{We also searched for archival data at 888 MHz 
from the Rapid ASKAP Continuum Survey \citep[RACS,][]{McConnell2020}. The first epoch RACS observations were 
performed between 2019 April 21 and 2020 June 21. We found 15 out of 18 sources were observed at 888 MHz, 
whose ASKAP images are shown in Appendix C.
We then measured the integrated and peak flux, following the same procedures described above, 
and found 13/15 sources are detected. 
For two non-detections, we report the 3$\sigma$ upper limit on the flux, based on the map rms at the off-source position, which 
is in the range {190--430 $\mu$Jy/beam}. 
Considering that the 3 GHz fluxes between the two epoch VLASS observations 
did not vary dramatically, the radio SED was likely evolving slowly. Therefore, the {RACS} data at 888 MHz can be considered 
quasi-simultaneously as the VLASS ones, which can be used to quantify radio spectral slopes below 3 GHz. 
We used the VLASS data that are close to the {RACS} observing date for each source to determine 
the rest-frame radio slope between 888 MHz and 3 GHz. 
The results are shown in Figure \ref{Fig:fitslope}. 
We found that the radio spectral {index ($\alpha_{\rm 0.89-3GHz}, S_{\nu}\propto\nu^{\alpha}$)}
is in the range 0.11--2.43, with a median of 0.87. 
With the radio spectral index, we extrapolated the 3 GHz flux to that at 1.4 GHz for 15 galaxies, 
and found it is a factor of $\simgt3-15$ higher than the FIRST flux, confirming the transient nature of radio emission. 
Details on the {RACS} flux measurements and derived radio spectral {index ($\alpha_{\rm 0.89-3GHz}$) }can be found in Table 2}. 

\begin{figure}
	\begin{minipage}[t]{0.41\linewidth}
		\centering
		\includegraphics[width=3.in]{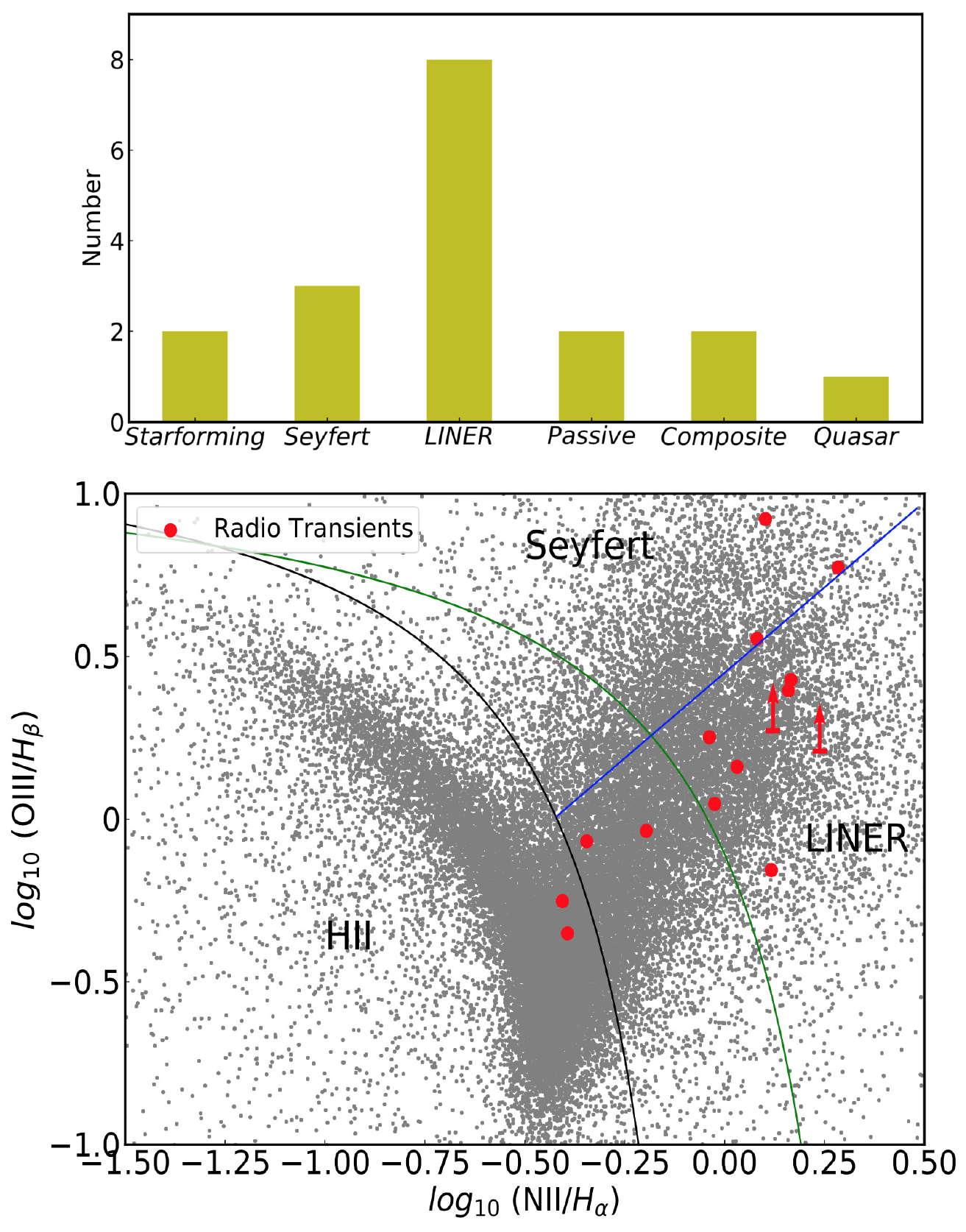}
	\end{minipage}
	\caption{The BPT diagram, based on the emission line ratios of {\sc [O iii]}/$H\beta$ vs. {\sc [N ii]}/$H\alpha$. 
	SDSS galaxies are shown in gray dots, while the radio transients in our sample are shown in filled red circles. 
	The different lines represent the boundaries commonly used to classify sources between star-forming galaxies, 
	LINERs, composite and AGNs, taken from \citet{Kewley2001, Kauffmann2003, Schawinski2007}. 
	The upper panel shows the classification of radio transients according to the emission line diagnosis of the nuclear activities. 	}
	\label{Fig:bpt}
\end{figure}

\begin{figure*}[bpt]
	\begin{minipage}[t]{0.48\linewidth}
		\centering
		\includegraphics[width=3.5in]{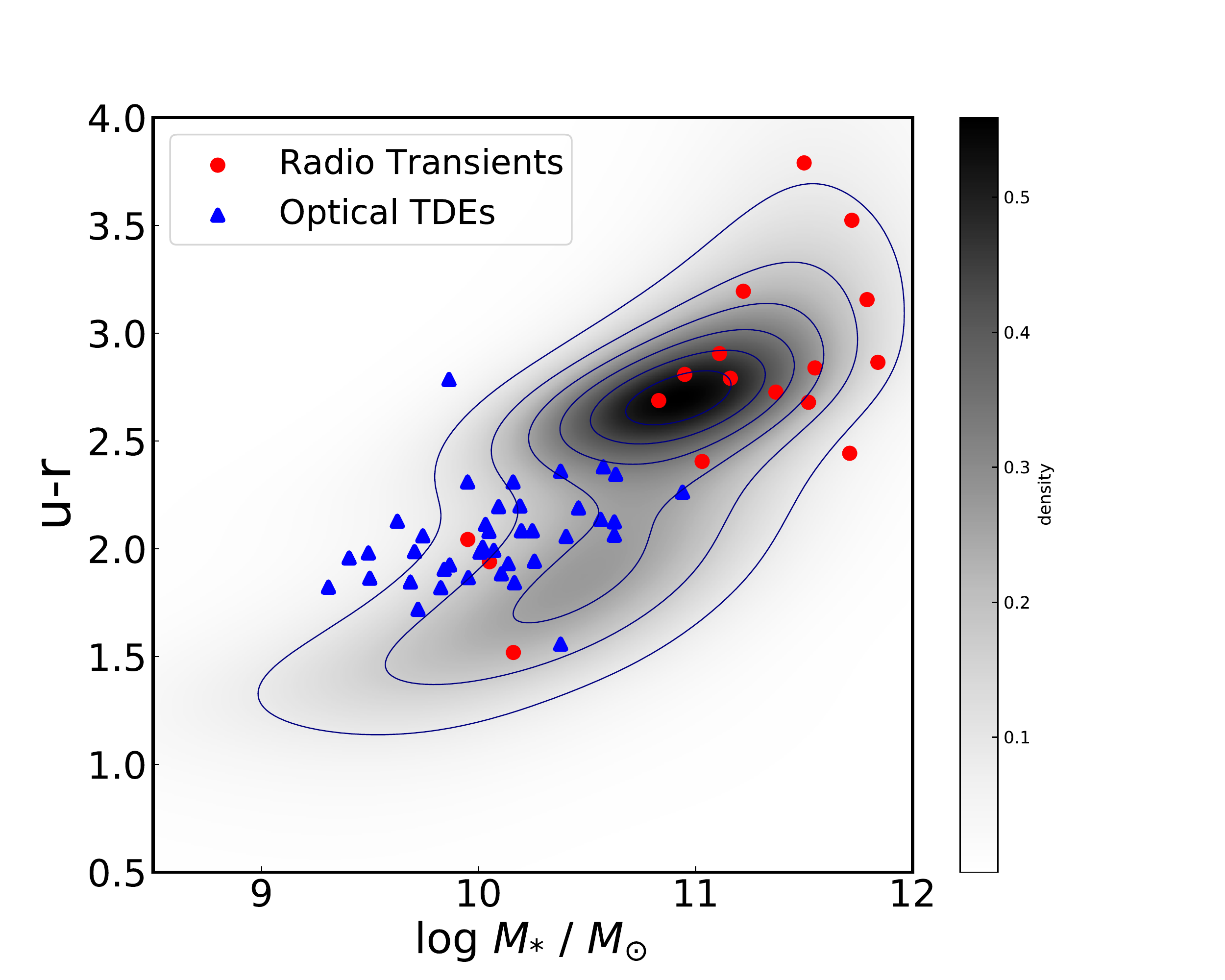}
	\end{minipage}
	\begin{minipage}[t]{0.48\linewidth}
		\centering
		\includegraphics[width=3.5in]{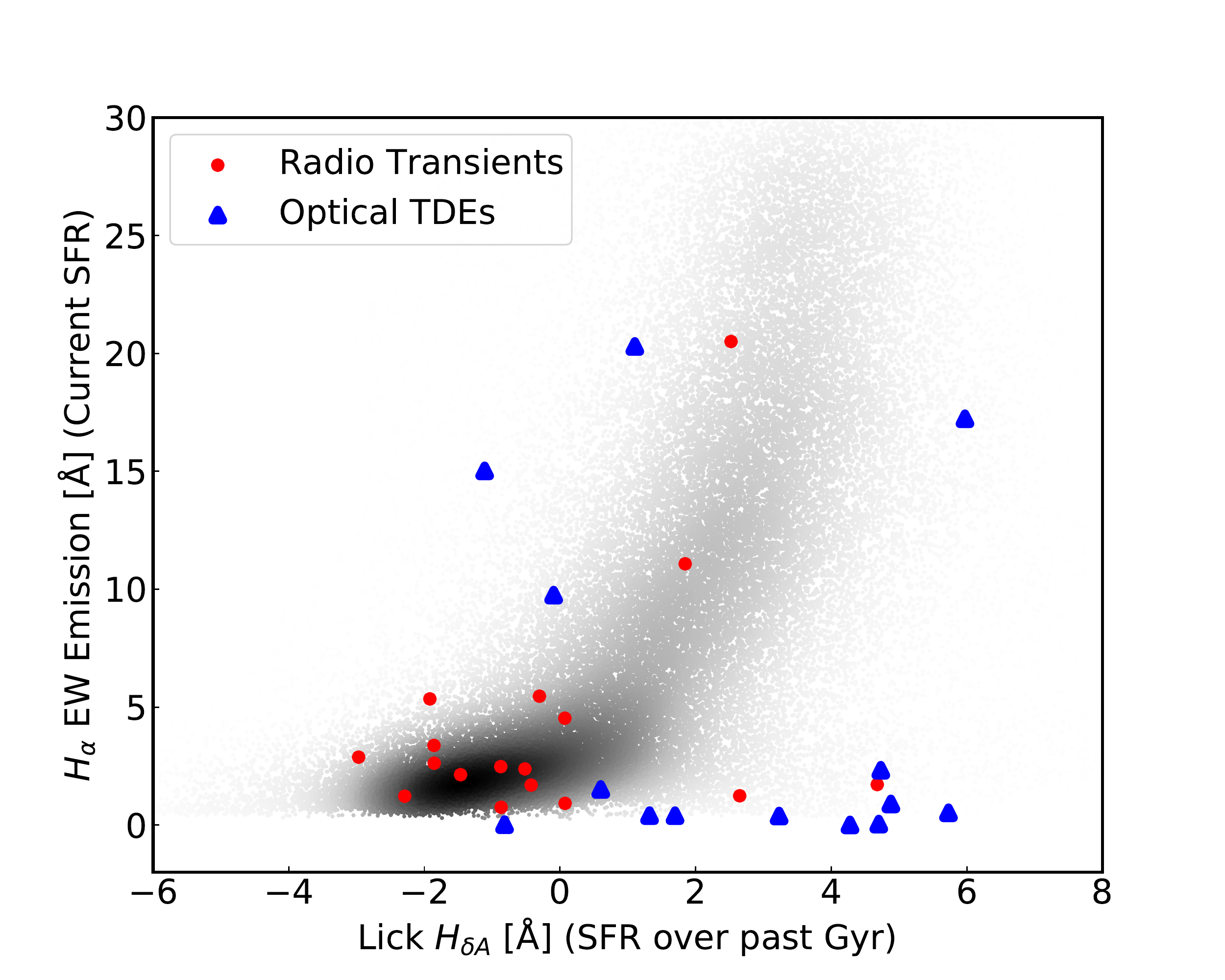}
	\end{minipage}
	\caption{{\it Left panel:} The extinction-corrected rest-frame $u-r$ colors versus total stellar masses. 
	Both are derived from the SED fittings to the combined SDSS and WISE photometry using MAGPHYS \citep{Chang2015}. 
	Contours show the density of SDSS spectroscopic galaxies, which display the bimodal distribution. 
	Filled red stars are our sample, with green stars for optical TDEs compiled in \citet{Velzen2021} for comparison. 
	{\it Right panel:}  Plot of the EWs H$\alpha$ emission lines versus the Lick H$\delta_{\rm A}$ indices. 
	SDSS galaxies are represented by gray dots. The radio transients in our sample and optical TDEs are marked with the same 
	symbols as left panel. }
	\label{Fig:cmd}
	 \vspace{0.5cm}
\end{figure*}

\begin{table*}
\centering
\caption{Jet power and host galaxy properties}
\setlength{\tabcolsep}{1.5mm}
\begin{tabular}{ccccccccccc}\hline\hline
Name & log$M_{\star}$ & log$M_{\rm BH}$ & $u-r$ & {Lick $H_{\delta A}$} & {EW(H$_{\alpha}$)} & log({\sc [N ii]}/$H_{\alpha}$) & log({\sc [O iii]}/$H_{\beta}$) & log $\lambda_{\rm EDD}$ & {log ($P_{\rm j}$/$L_{\rm bol}$)} & Type\\
 & ($M_{\sun}$) & ($M_{\sun}$) &  & [Å] & [Å] & & & & & \\\hline
J0040+0823 & 11.84 & 9.086 & 2.865  & 2.649 & 1.237 & 0.121 & $>$ 0.272 & $>$ -3.71 & $<$ -0.61 & LINER\\
J0154-0111 & 10.83 & 8.761 & 2.687  & -1.858 & 3.376 & 0.117 & -0.157 & -3.97 & -1.10 & LINER\\
J0800+2928 & 10.05 & 5.619 & 1.940  & 4.678 & 1.717 & 0.102 & 0.922 & -0.51 & -1.24 & Seyfert \\
J0951+3703 & 11.50 & 8.500 & 3.790  & -1.919 & 5.349 & -0.345 & -0.068 & -2.34 & $>$ -1.75 & Composite\\
J1029+0436 & 10.95 & 8.035 & 2.809  & -2.970 & 2.879 & -0.025 & 0.047 & -3.05 & -0.63 & LINER\\
J1129+3900 & 11.72 & 8.277 & 3.524  & -0.873 & 2.478 & 0.166 & 0.428 & -1.25 & -2.21  & LINER\\
J1217+2750 & 11.16 & 8.026 & 2.791  & -1.466 & 2.136 & $\sbond$ & $\sbond$ & $\sbond$ & $\sbond$  & Passive\\
J1301+2127 & 10.16 & 6.098 & 1.519  & 2.522 & 20.505 & -0.406 & -0.252 & $<$ -0.89 & $>$ -1.14 & Starforming\\
J1337+3857 & 11.71 & 8.253 & 2.443  & 0.075 & 0.919 & $\sbond$ & $\sbond$ & $\sbond$ & $\sbond$ & Passive\\
J1407+1247 & 11.37 &  7.671 & 2.727  & -2.290 & 1.223 & 0.160 & 0.396 & -1.76 & -1.57 & LINER\\
J1409+5420 & 11.11 &  7.893 & 2.906  & -1.855 & 2.624 & 0.023 & $>$ 0.208 & $>$ -2.69 & $<$ -$0.70^{\dag}$ & LINER\\
J1437-0033 & 11.79 &  8.418 & 3.156  & -0.868 & 0.752 & 0.031 & 0.161 & -4.20 & 0.43 & LINER\\
J1558+1412 & 11.03 &  7.842 & 2.405  & -0.517 & 2.381 & -0.038 & 0.252 & -2.62 & -1.55 & LINER\\
J1610+0606 & 11.22 &  7.724 & 3.195  & -0.425 & 1.693 & 0.081 & 0.554 & -1.29 & $>$ -2.06 & Seyfert\\
J1642+3346 & 11.52 &  7.618 & 2.680  & -0.303 & 5.463 & -0.196 & -0.036 & -1.59 & -1.63 & Composite\\
J1646+4227 & 9.95  &  6.415 & 2.043  & 1.846 & 11.077 & -0.393 & -0.351 & $<$ -1.38 & $>$ -$1.25^{\dag}$ & Starforming\\
J2301+0544 & 11.55 & 8.433  & 2.839  & 0.072 & 4.526 & 0.284 & 0.773 & -0.77 & -3.16 &  Seyfert\\\hline
\end{tabular}
{{Note- $^{\dag}$The two sources are not covered by ASKAP observations, so we used the median value of $\alpha_{\rm 0.89-3 GHz}=0.87$ 
 to derive the rest-frame 1.4 GHz flux hence {jet power \citep[e.g., Equation (4),][see also Section 5.4]{Wolowska2021}.  }}}
\end{table*}

\section{Results} \label{sec:Results}
\subsection{Host galaxy properties}
In order to understand the nature of the VLASS RTS, it is useful to assess the nuclear activity of these galaxies prior to the {radio brightening detected by VLASS}. 
Along with the SDSS data release, there are several works to provide value-added catalogs of 
the galaxy intrinsic properties\footnote{https://www.sdss.org/dr12/spectro/galaxy\_portsmouth, and 
https://www.sdss.org/dr12/spectro/galaxy\_mpajhu}.  
We used the MPA-JHU SDSS spectroscopic catalog for our following analysis. 
After cross-matching with the catalog, we found that there are 15 out of 18 galaxies with 
emission-line flux measurements (or upper limits). The remaining three are not listed in the MPA-JHU 
catalog, likely because the emission lines are too weak to be detected. We carried out 
detailed spectral fittings to measure any emission lines if present, and the results are shown 
in {Appendix D}. Two sources, J1217+2750 and J1337+3857, have no detectable 
emission lines, while only upper limits can be obtained on the H$\beta$ lines for J0040+0823 and J1409+5420. 
The line ratios of  {\sc [N ii]}/$H\alpha$ and {\sc [O iii]}/$H\beta$ are listed in {Table 3}. 
Note that one object, J0950+5128, shows clearly the spectral features of quasars {(Figure A1 in Appendix A),} 
{and is listed in the quasar sample that have transitioned from radio-quiet to radio-loud \citep{Nyland2020}. 
Since the optical spectrum of J0950+5128 is dominated by quasar emission, making it a challenge to 
study the host galaxy and its stellar population, we do not consider it in our following analysis relevant to the 
host properties {(Figures 4, 7 and 9)}. }
Using the emission-line flux ratios, we performed a diagnosis of the nuclear activity by 
classifying the sample into different subclasses according to their location on the
BPT diagram \citep{Kewley2001}, as shown in Figure \ref{Fig:bpt} (lower panel). 
Among 15 galaxies with emission-line measurements, three are classified as Seyfert 2 
galaxies, two as star-forming galaxies, {eight} as low-ionization nuclear emission regions (LINERs), 
and two as composites. 
For the remaining two objects with non-detections of emission lines, we consider
them to be passive galaxies. 
It should be noted that a fraction of LINERs may actually be ionized by
evolved stars or radiative shocks rather than by AGNs, as LINER-like
spectra often appear in off-nucleus regions of normal galaxies
as well \citep[e.g.,][]{Cid2011, Zhang2017}. 
In summary, in addition to one quasar, only 3 out of 17 galaxies in our sample ($\sim$16\%) display 
Seyfert-like narrow emission-line ratios, indicating that most, if not all, have no strong  
nuclear activities in the {SDSS} optical spectra (Figure \ref{Fig:bpt}, upper panel). 
{Since the narrow emission-lines are expected to originate from regions at $\sim$kpc scales, 
the AGN classification in principle is not biased against obscured type 2 sources \citep[e.g.,][]{Kauffmann2003}. }

Having established that most of the {RTS} in our sample have a weak AGN component, 
we now investigate the host galaxy properties {(excluding J0950+5128)} in order to further understand the origin of the radio emission. 
The MPA-JHU catalog has also provided the measurements of stellar mass ($M_{\star}$) through SED fitting 
with stellar population models. 
However, their fittings only used the SDSS optical photometry (up to $\sim$9000\AA), 
and may introduce bias in estimating stellar masses. 
 Hence, we used the publicly released catalog with the SED fitting results by combining the SDSS and WISE photometry \citep{Chang2015}, 
 covering a broader wavelength range of $\lambda=0.4-22\mu$m. 
We calculated the $u-r$ colors based on the rest-frame $u$-band and $g$-band luminosity, for which 
the dust extinction and $k-$correction have been taken into account from the best-fit SED for each galaxy.  
For three galaxies that are not presented in the catalog of \citet{Chang2015}, 
we derived their host stellar masses by fitting the SDSS and WISE photometry using the CIGALE code (Appendix D). 
It is well known that galaxies show a bimodal distribution in the color-magnitude diagram, which
are mainly divided into a red sequence and blue cloud according to the colors, with a green valley in between \citep[e.g.,][]{Strateva2001, Bell2004}. 
Figure \ref{Fig:cmd} (left) shows the extinction-corrected, rest-frame $u-r$ color versus total stellar mass for 
our galaxies ({filled red circles}). 
The SDSS spectroscopic sample from \citet{Chang2015} are overlaid in contours for comparison. 
It can be seen that the RTS host galaxies are dominated by the red and massive galaxies, 
with the majority (14/17) falling into the locus of red sequence. 
\begin{figure*}[t!]
		\centering
		\includegraphics[scale=0.8]{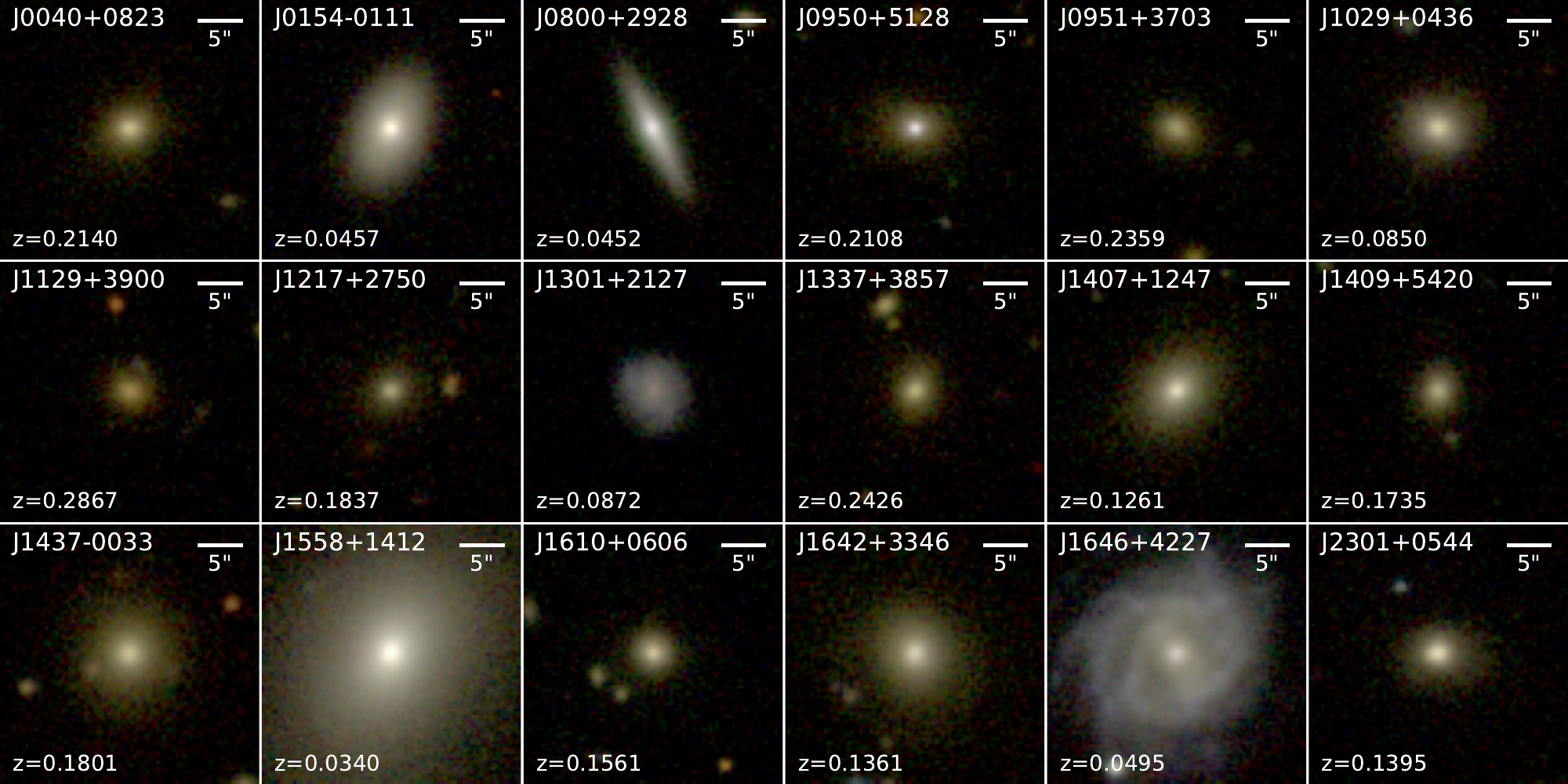}
	\caption{Composite SDSS {\tt gri} color images of the RTS host galaxies.  
	Each panel has a size of 30\arcsec$\times$30\arcsec. 
	}
	\label{Fig:morph}
	 \vspace{0.5cm}
\end{figure*}


Since the colors of galaxies are affected by both the current star-formation rate (SFR) and 
star formation history, {it has been proposed that the H$\alpha$ equivalent width (EW) and Lick 
H$\delta_{\rm A}$ index are useful to quantify the host properties according to the star-formation history \citep{French2016}. 
The H$\alpha$ EW can be used as a tracer for the current star-formation history on timescales of $\sim$10 Myr, 
while the H$\delta_{\rm A}$ index can probe star-formation on much longer timescales of $\sim$1 Gyr.  }
Figure \ref{Fig:cmd} (right) shows the H$\alpha$ EW versus Lick H$\delta_{\rm A}$ index for SDSS galaxies 
from the MPA-JHU catalog, {and RTS in our sample}. 
Apparently, the majority sources in our sample  
(13/17) fall into the region with little-to-no H$\alpha$ emission and low H$\delta_{\rm A}$. 
To encompass the 13 sources, we defined the selection cut H$\delta_{\rm A}\simlt0$ and H$\alpha$ EW$<7$\AA. 
This cut includes 242249 SDSS galaxies, or a fraction of $\sim$32\%. 
{This indicates that the host galaxies for the majority of RTS are characterized by 
low recent specific star formation rate with relatively old stellar population, consistent with 
normal massive galaxies in the red sequence.  
In comparison, the host galaxies of optically-selected TDEs (filled blue triangles in Figure \ref{Fig:cmd}) 
appears to show different stellar properties, and we will discuss the implications in detail in Section 5.2. 
}

{In Figure \ref{Fig:morph} we show cut-outs of the SDSS color {\tt gri} images for the 18 RTS galaxy hosts. 
{We include the quasar J0950+5128 for the morphology analysis, as it is resolved in the SDSS imaging.}
 Galaxies can be classified as elliptical-like or disk-like based on the concentration index $C=R_{90}/R_{50}$, 
 where $R_{90}$ and $R_{50}$ is the radii containing 90\% and 50\% of the Petrosian flux for a given band, respectively, 
 and the likelihood ratio of the de Vaucouleurs' model fit to that of the exponential model.  
 Elliptical-like early-type galaxies can be selected with $C>2.5$ in $i-$band, and the likelihood ratio $>$1.03 \citep{Bernardi2003}. 
 Using the measurements derived with the SDSS pipeline, we found 16 out of 18 galaxies ($\sim$88\%) meet the criterion $C>2.5$ for  
 early-type galaxies, which can also be better fitted by a de Vaucouleurs profile. 
 On the other hand,  we cross-matched with the Galaxy Zoo DR1 catalog \citep{Lintott2011} and found 14 matched galaxies. 
 Among the 14 galaxies, 10 are more likely elliptical (debiased probability $P>0.7$) based on visual classifications, 
 two are spiral galaxies and two are ambiguous. 
 Thus, we conclude that more than 70\% of the sample are early-type galaxies, 
 consistent with the results of stellar population analysis. 
 
}
\begin{figure*}[htb]
\centering{
 \includegraphics[scale=1]{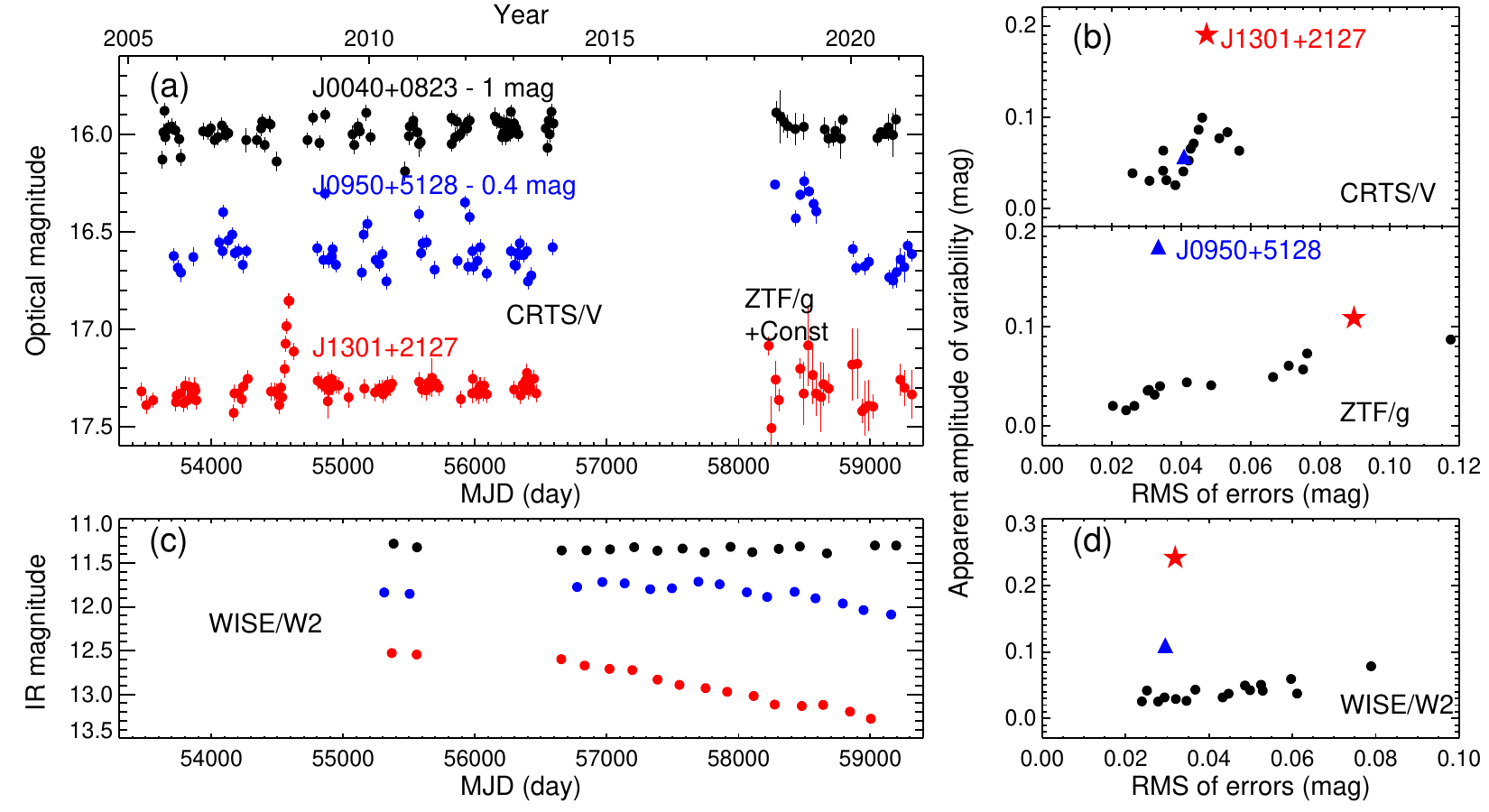}
 \caption{:
  Optical and MIR time-domain data of our sample and the analysis. (a) CRTS V-band and ZTF g-band LCs of three typical galaxies: J0040+0823 (black) represents those without optical variability, while J0950+5128 (blue) and J1301+2127 (red) are two variable galaxies. For clarity, we display the monthly-binned ZTF LCs by adding constants. (b) The apparent variability amplitude versus the RMS of the photometric errors for the CRTS V-band and ZTF 
  g-band observations. (c) WISE $W_2$-band LCs of the three galaxies. (d) Similar to (b), but WISE/$W_2$ data were used.
  }}
\label{fig_optmir}
 \vspace{0.5cm}
\end{figure*}

\subsection{Variability properties at optical and MIR bands}
{We now check if there are optical flares associated with these radio transients.}
We collected optical light-curve (LC) data from Catalina Real-time Transient Survey \citep[CRTS] {Drake2009} and Zwicky Transient Facility \citep[ZTF] {Bellm2019}.
{Among the 18 objects in the sample, 16 were observed by both CRTS and ZTF, and one (J1029+0436) was only observed by CRTS.
The remaining one (J1646+4227) was not observed by either CRTS or ZTF. }
For the CRTS data, the V-band LCs observed from 2005 to 2013 are public\footnote{Data can be obtained from Catalina survey data releases 2 (http://nesssi.cacr.caltech.edu/DataRelease/)}. And for ZTF, the LCs in g, r and i bands from 2018 are public\footnote{Data can be obtained from NASA/IPAC InfRared Science Archive (https://irsa.ipac.caltech.edu/applications/Gator/)}.
{As can be seen in Figure \ref{fig_optmir}(a),} the quasar J0950+5128 shows stochastic variability with amplitude of $\sim$0.2 magnitude, which is consistent with typical variability amplitude found in AGN \citep{Caplar2017}.
By visually inspecting the optical LCs, we found that {for most of the non-quasar galaxies, the LCs are roughly consistent with flat lines with noise, and do not show flares with amplitude more than 0.2 magnitude.}
In Figure \ref{fig_optmir}(a) we show the LC of the source J0040+0823 as an example.
The only non-quasar galaxy that shows obvious variability is J1301+2127, whose LC is also displayed in Figure \ref{fig_optmir}(a). J1301+2127 shows a flare between April and July 2008. The peak of the flare was recorded in May 2008, when the galaxy was brightened by 0.5 magnitude compared to the pre-flare level. In order to check the reliability of the variations, we calculated the ``apparent'' amplitude of variability for our sample, and compared it with the typical error level of optical photometry. Here the ``apparent'' amplitude of variability means that any variations caused by photometric errors has not been removed. For the ZTF data, the amplitude of variability is calculated as the standard deviation of the magnitudes from the monthly-binned g-band LC. For the CRTS data, it is calculated similarly in a two-years-long period with the greatest amplitude. The typical error level is estimated using the root mean square (RMS) value of the photometric errors. As shown in Figure \ref{fig_optmir}(b), for most non-quasar galaxies the amplitudes of variability {are comparable or less than the typical errors, except for J1301+2127}. The analysis supports the results from visual inspection. J1301+2127 is classified as a star-forming galaxy in the preflare optical spectrum. The optical flare in J1301+2127 lasts for several months, which is similar to those of many TDEs \citep{Velzen2021}. 
{In summary, among the 18 galaxies with transient radio emission, only one shows unambiguous evidence of optical flare. 
}

{It is possible that optical flares (if present) could be missed occasionally due to the low cadence of optical photometric observations.
We estimate the probability that a hypothetical TDE causing the radio flare is captured by the optical photometric observations.
The observational cadence of the CRTS is: during the 9-year period from 2005 to 2013 (8 years for a few sources), monitoring lasted for 4--9 months in each year (the average value is 5.8 months for our sample); in each month there are 1--3 observing nights; in each night there are 4 exposures. 
By combining the data taken from the 4 exposures, the photometric error is 0.03--0.06 magnitude for our sample. 
As a TDE typically causes a brightening with 0.2--1 magnitude near the peak time \citep{Velzen2021}, we estimated that a TDE would 
be detected as long as there are CRTS observations in the month when the TDE peaks. 
There are 15 non-quasar galaxies in our sample that were observed by CRTS. 
For each galaxy, there are 51 months on average covered by CRTS observations. 
If assuming that the TDE peak time is uniformly distributed between the FIRST and the epoch I VLASS observations, 
which have a time interval of 249 months on average, the probability of a TDE being captured by CRTS is 20.5\%. 
The expected number of optical flares detected by CRTS in these 15 galaxies can be estimated to be 3.1. 
For the ZTF survey the observation cadence is: starting from March 2018, the monitoring lasted for 9--10 months in each year; 
in each month, there were 10--30 observational nights; and in each night there was one exposure in each band. 
The photometric error of each observation is 0.02--0.12 magnitude for our sample. 
Thus the cadence of single observation or observations in several consecutive nights is sufficient to detect a TDE peak lasting for one month. 
We then calculated the probability that the TDE peak is captured by ZTF survey as we did for the CRTS observations. 
There are 7 non-quasar galaxies in our sample which were observed 
during the period between ZTF and epoch I VLASS observations.  
The summed duration of the ZTF observations is 11 months on average. 
Thus the probability of a TDE being detected is 4.4\% (11/249), and the expected number in these 7 galaxies is 0.3. 
By adding the detection probability in the CRTS and ZTF light curves, the expected number of optical flares 
recorded by the two surveys is 3.4. Note that the number is likely to be underestimated. 
This is because the optical flare of a TDE typically lasts for several months, and there is a possibility that its rising or falling phase can be recorded although the peak time is missed. 
In addition, if the radio emission lasts for less than 10 years, as expected for many optical TDEs \citep[e.g.,][]{Alexander2020}, 
the TDE occurrence time can be better constrained to be after 2004 for $z$$<$0.3, a period with more overlap with the optical photometric observations. 
This will also increase the probability of TDE flares in our sample being detected by optical surveys. 
However, only one optical flare (J1301+2127) is actually detected. 
Therefore, the low detection rate of optical flares cannot simply be explained by the low cadence of the CRTS or ZTF observations. 
}

An optical flare may also be missed due to dust obscuration. Fortunately, dust heated by a central UV/optical flare can reradiate in the infrared (IR), {causing 
an IR flare as well as echo emission \citep[e.g.,][]{Mattila2018, Jiang2021a}. 
The IR flares caused by TDEs usually lasts for 1--10 years with amplitudes of 0.2--2 magnitudes \citep[e.g.,][]{Jiang2016, Jiang2017, Jiang2019, Jiang2021b, Dou2016, Dou2017}.}
We searched for IR flares using the IR data in the WISE $W_2$ band (at 4.6 $\mu$m). In a way similar to what we did for optical data, {we calculated the variability amplitudes as well as the typical errors,} as shown in Figure \ref{fig_optmir}(c) and (d). {For most of non-quasar object the amplitudes of variability are comparable or less than the errors, and the only galaxy with apparent IR variability amplitude larger than 0.2 magnitude is J1301+2127. }
The long decaying IR emission in J1301+2127 is similar to those of TDE candidates with extreme coronal emission lines \citep{Dou2016}.
{For those galaxies without an optical flare, there is also no significant IR variation as the amplitudes of variability are less than 0.1 magnitude.
Note that the cadence of WISE observations is once every six months and the photometric errors are 0.02--0.08 magnitudes, 
which are sufficient to detect the IR echoes if it present.
Therefore, we did not find evidence of the dust-heated IR echoes from central optical flares for most our galaxies.}


\section{Discussion} \label{sec:Discussion}

We have identified 18 galaxies at $z<0.3$ that display {transient} radio emission by comparing between the radio flux 
observed with VLASS and the upper limit with FIRST. It should be noted that the sample, by selection, is only 
sensitive to the long-term radio variability properties on time scales of years to decades, and the cadence of 
observations is very sparse. 
{We have carefully inspected the radio components detected in the epoch I VLASS observations, and 
found that they are not spurious sources. The result is supported by the detections in the epoch II VLASS observations 
for all sources in the sample. }
Long-term radio variabilities could have different physical processes, such as a supernova (SNe) 
explosion, intrinsic variability of a {radio-quiet AGN}, sporadic accretion onto a SMBH due to instability in an accretion 
disk or tidal disruption of a star, or a recently launched young jet. We will discuss these scenarios in detail.

\subsection{Stellar explosions}
Radio emission has been detected in about 30 per cent of the nearby SNe sample ($D<100$ Mpc) at a few days 
to several years since the explosion \citep{Bietenholz2021}. 
A recent study on the luminous late-time radio emission from SNe detected by VLASS has been presented in \citet{Stroh2021}. 
{As described in Section 3, we have cross-correlated with the open SNe catalogue compiled by \citet{Guillochon2017} 
and found two matches in our initial sample of 20 galaxies.  
}
After excluding the two known SNe, the radio luminosities of our sources are high, with $L_{\rm 3GHz}\simgt10^{39}$ erg $s^{-1}$, 
which are comparable to the most luminous Type IIn SNe \citep{Stroh2021}, for which the host galaxies are blue with active ongoing star-formation. 
This is not compatible with the fact that most of our galaxies have red colours and current star-formation rates 
are low. The high radio luminosities rule out the possibility of SNe for the {transient radio emission}.
The contribution to the observed radio variability by gamma-ray bursts (GRBs) seems also unlikely, as 
typical variability timescale of radio emission for GRBs is around 1-2 weeks \citep{Pietka2015}.  
Long GRBs can have a radio luminosity as high as $10^{40}-10^{41}$ erg $s^{-1}$ \citep{Stroh2021}, 
but decay {by more than an order of magnitude within several years since explosion, which is inconsistent 
with lack of significant flux variability of our sources between two-epoch VLASS observations ($\simlt$50\%, Figure \ref{Fig:fluxvar}). }
Furthermore, the effect of interstellar scintillation causing the radio {variability} 
can be ruled out either, as the 
flux variability due to interstellar scintillation is at a level of $\sim$30\%--40\% \citep{Nyland2020}, 
much lower than {the radio brightening by a factor of $>$300\% between FIRST and VLASS observations (Section 3)}. 

\subsection{Tidal Disruption Events}

Recent time-domain surveys have discovered dozens of TDEs in the centers of otherwise quiescent 
galaxies \citep{Gezari2021}, but only $\sim$10\% are accompanied by radio flares \citep{Bower2013, Velzen2013, Alexander2020}, the nature of which is still poorly understood \citep{Horesh2021a, Horesh2021b}.  
So far there are two TDE candidates that have been solely identified at the radio wavelengths \citep{Anderson2020, Ravi2021}. 
{It is possible that some of the {RTS} in our sample are associated with TDEs. 
However, as shown in Figure \ref{Fig:cmd}, we found the red colors and large host stellar masses for the host galaxies of RTS.  
Based on the TDE sample selected from the ZTF survey and known UV/optical-selected 
TDEs in literature, \citet{Velzen2021} 
show that the TDE host galaxies are preferred to locate within the green valley region, 
where the galaxies are expected to be transitioning from star-forming to quiescent. 
A Kolmogorov–Smirnov (K-S) test for the stellar mass distribution between 
RTS and optical TDEs results in a $p-$value of $3.3\times10^{-8}$, suggesting that they are not drawn from the same population\footnote{
The estimate on the true $p-$value might be affected due to the small sample sizes used.}.   
A similar result can be obtained from the K-S test on the distribution in the rest-frame $u-r$ colors, with a $p-$value of $3\times10^{-8}$.  
It has been found that UV/optical-selected TDEs are over-represented by post-starburst galaxies \footnote{The central concentration 
of A stars in post-starburst galaxies could serve to enhance the TDE rate \citep{Yang2008, Stone2016}.}.  
Figure \ref{Fig:cmd} (right) shows that only two RTS lie in the branch for the TDE hosts (quiescent Balmer-strong galaxies).  
There are two sources having a Lick H$\delta_{\rm A}$ index consistent with TDE hosts but with enhanced H$\alpha$ emission, possibly 
dominated by the current star-forming activity and/or AGN. 
The above analyses suggest 
that the host properties of RTS presented in this work may be different from that of TDEs, at least for those 
discovered in the optical surveys.

The red colors and large stellar masses suggest that RTS 
may possess larger SMBHs than known TDEs or TDE candidates.  }
By using the velocity dispersion measurements provided in the MPA-JHU catalog,  and 
adopting the $M_{\rm BH}-\sigma_{\star}$ relation used in \citet{Gultekin2009}, we estimated  
their BH masses in the range $4.2\times10^{5}-1.2\times10^{9}$\msun, with a median of $1.1\times10^{8}$\msun. 
Figure \ref{Fig:bh}  shows the distribution of BH masses, and its comparison with the sample of 12 
optical/UV selected TDEs with the BH masses measured by \citet{Wevers2017}.  
{It can be seen that our sample is much more spread out, with objects at either large or small masses, 
indicating the heterogeneity of the sample. 
A Kolmogorov-Smirnov (K-S) test on the mass distribution between our sample and optical TDEs 
results in a p-value of 2.9$\times10^{-5}$, suggesting that they are not drawn from the same population.
}
While 10 of 12 optical TDEs have masses between $3\times10^{5}$ and $10^{7}$\msun, 
only {3 out of 17} sources in our sample fall into this range. 
Among the three sources, two can be classified as star-forming galaxies, namely J1301+2127 and J1646+4227. 
As shown in Section 4.2 (and Figure \ref{fig_optmir}), the clear optical flare and long-lasting IR echo suggest that J1301+2127 
might be a TDE. 
{J1646+4227 has no IR flux variations detected in the WISE light curves. 
The optical variability properties are not clear yet, as it is the only source in our sample without CRTS or ZTF observations. 
We note that very few optically-discovered TDEs are found in star-forming galaxies, and 
this could be a selection effect \citep{Jiang2021b, Jiang2021a}. 
If J1301+2127 and J1646+4227 were explained due to TDEs, 
it would suggest that radio observations are important to offer a complete view of TDE phenomenon.
}
On the other hand, {more than half sources in our sample (9/17=53\%)} 
have BH masses larger than $10^{8}$\msun, which is the upper limit 
for tidal disruption of a main-sequence star around a Schwarzschild black hole. 
In the TDE scenario, this suggests either the disruption of a post main-sequence star 
\citep{Kochanek2016} or the black hole is spinning \citep{Leloudas2016}. 
{However, such events are expected to be rare based on the current observations of TDE candidates. 
In addition, it is difficult to reconcile with the non-detections of optical and MIR flares in the historical CRTS, ZTF and WISE data sets. 
The analysis of long-term optical and MIR light curves, host properties and BH mass distribution  
suggest that most of RTS in our sample are likely not TDEs. }


\begin{figure}[tb]
	\begin{minipage}[t]{0.48\linewidth}
		\centering
		\includegraphics[width=3.3in]{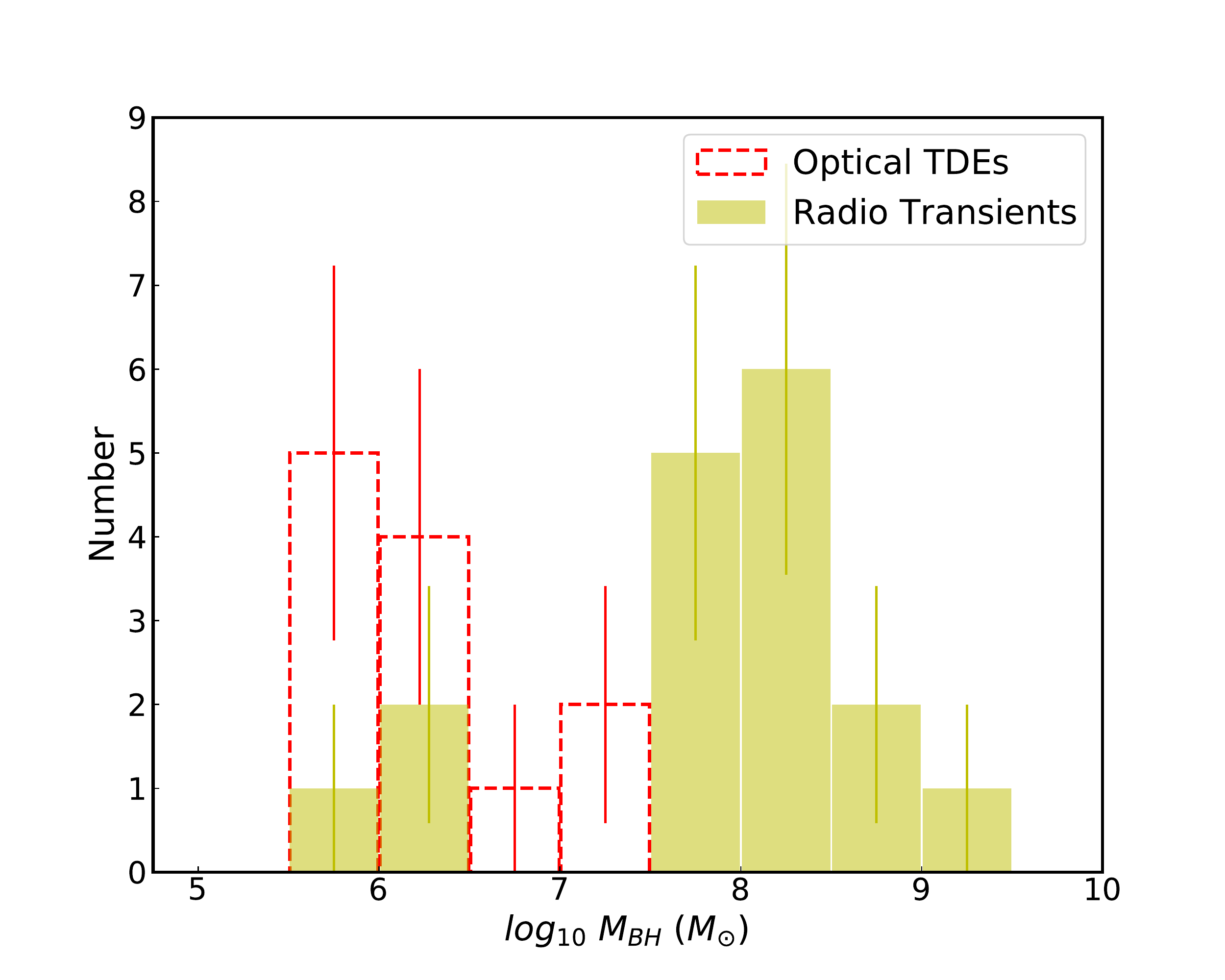}
	\end{minipage}
	\caption{
	A comparison of BH mass distributions for our sample (dark yellow) and optical/UV TDE galaxies 
	from \citet{Wevers2017}. Both BH masses are derived using the stellar velocity dispersions. 
	{Error bars represent 1$\sigma$ Poisson uncertainties, i.e., $\sigma_{\rm k}=\sqrt{N_{\rm k}}$ where $N_{\rm k}$ 
	is the number of measurements falling in $k$-th bin of $M_{\rm BH}$.  }
	}
	\label{Fig:bh}  
\end{figure}

\subsection{Intrinsic Variability of Radio-quiet AGN}

{As shown in Section 4.1, 4 out of 18 galaxies in our sample can be classified as 
AGNs (Seyfert or quasar) based on optical spectroscopy characteristics. 
However, the four AGNs are expected to be radio-quiet during the period of FIRST observations. 
In fact, all our sources have a low radio luminosity ($L_{\rm 1.4 GHz}\simlt10^{21}-10^{23}$ W Hz$^{-1}$) 
based on the weak or non-detections in the FIRST survey. 
This is at least an order of magnitude lower than the luminosity thresholds of $10^{24}-10^{25}$ W Hz$^{-1}$ 
for radio-loud AGNs \citep{Miller1990, Goldschmidt1999}. 
While radio variability has been observed in radio-quiet AGNs over months to years time-scales, 
the variability amplitude is typically a few tens per cent \citep{Panessa2019}. 
This contrasts with the large flux increases by a factor of $\simgt$3--15 between the FIRST and VLASS observations. 
Considering that the majority of our sources (14/18) can be classified as LINERs or normal galaxies, 
the intrinsic AGN variability seems unable to explain the radio properties of our sources. 

}

\begin{figure}[t!]
 \begin{center}
 \includegraphics [width=0.45\textwidth]{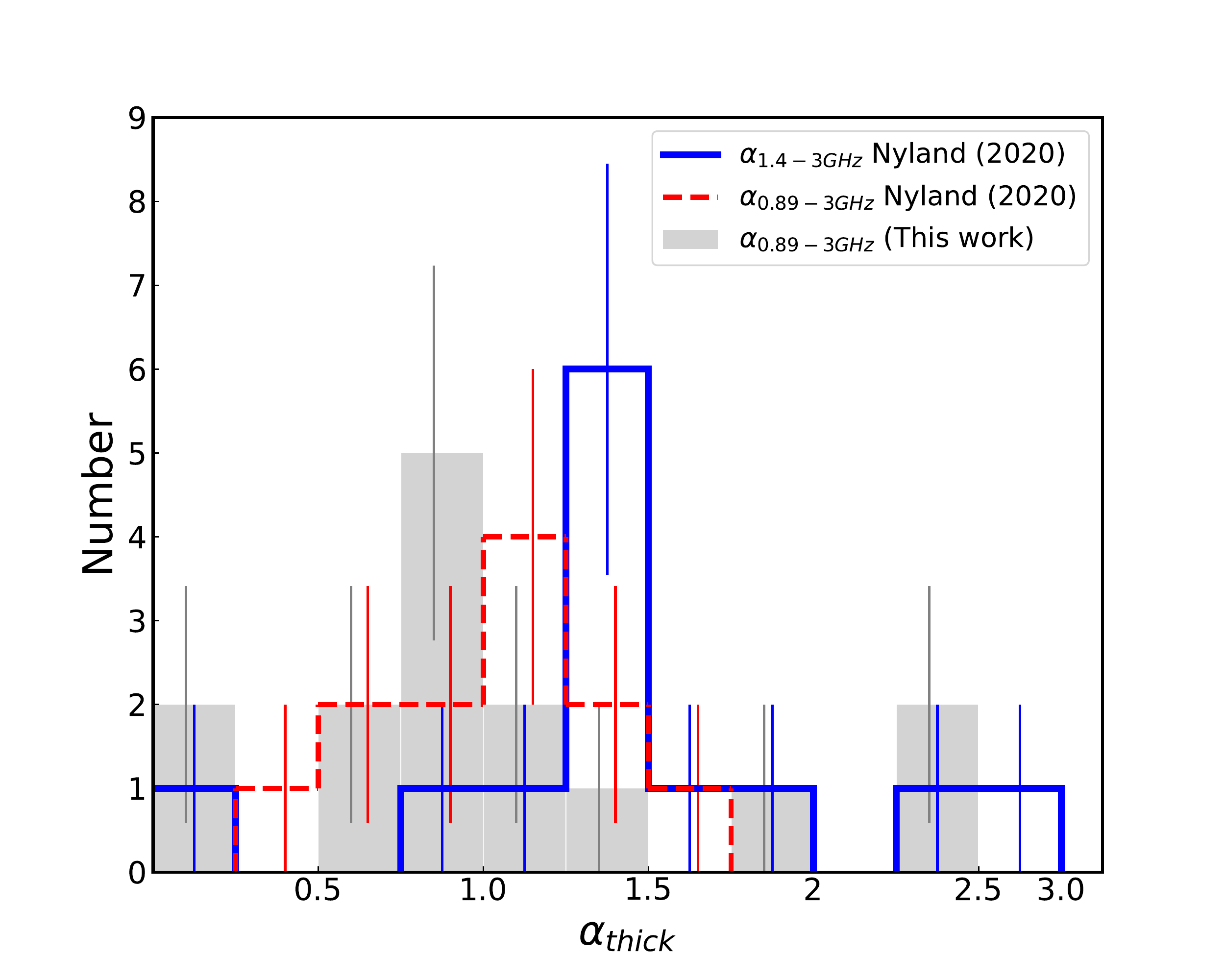}
 \hspace{1.5cm}
 \end{center}
 \vspace{-0.2cm}
 \caption{Histogram of the radio spectral index between 888 MHz and 3 GHz for RTS in our sample and those 
 in \citet{Nyland2020}. For comparison, we also show the optically-thick spectral indices ($\alpha_{\rm thick}$) 
 for the latter sample, which were derived by fitting a power-law model to the quasi-simultaneous flux densities at 
 1.4 GHz and 3 GHz. 
 {Error bars represent 1$\sigma$ Poisson uncertainties.}
 }
 \vspace{0.2cm}
 	\label{Fig:slope}  
 \end{figure}

\subsection{Recently Launched Young Radio Jets}
We next consider the possibility that the central black hole is in a long quiescent level, 
and fed episodically for a relatively short time-scale, probably caused by 
instability within an accretion disk. The low-level accretion activity in the quiescent state 
may also explain the very common LINER optical spectra for most of our sources (Section 4.1). 
It has been hypothesized that the radio activity is an intermittent phenomenon lasting 
$10^{4}-10^{5}$ year, and an AGN may undergo many such short-term phases during its lifetime 
\citep[e.g.,][]{Reynolds1997, Czerny2009, An2012, Wolowska2017}. 
Using the CNSS and VLASS data, recent works have identified a population of quasars with switched-on radio activities 
\citep{Mooley2016, Kunert2020, Nyland2020, Wolowska2021}. 
The transition from radio-quiet to radio-loud phase in these sources can be explained by the appearance of new-born and  
young radio jets. 

{Young radio jets are characterized by curved radio SEDs peaking at $\sim$5--10 GHz \citep{Nyland2020}. 
The spectral turnover at frequencies below the peak is likely due to the synchrotron self-absorption and/or free-free 
absorption \citep{Odea2021}. 
Figure \ref{Fig:slope} shows the distribution of the radio spectral index between 888 MHz and 3 GHz for our sources. 
Interestingly, all have a flat or inverted spectrum, with $\alpha_{\rm 0.89-3 GHz}$ in the range 0.1--2.4. 
For comparison, we also show the optically thick spectral index values ($\alpha_{\rm thick}$) below the turnover 
frequency for the sample presented in \citet{Nyland2020}, which appear to be higher than ours.  
A K-S test for the two samples results in a $p-$value of 0.01. 
Such a difference could be due to the poor spatial resolution of the ASKAP observations ($\sim$15\arcsec), 
which may contain the radio emission at a larger scale than the VLASS one. 
In addition, $\alpha_{\rm thick}$ for the latter sample was derived between 1.4 GHz and 3 GHz, which 
has a different frequency coverage. 
Therefore, we retrieved the ASKAP data at 888 MHz and measured the $\alpha_{\rm 0.89-3 GHz}$ for the 
sample of \citet{Nyland2020}, whose distribution is also shown in Figure \ref{Fig:slope}. 
In this case, a K-S test ($p=0.65$) suggests that it could be drawn the same population as ours. 
Although precise measurements of radio SEDs can not be done with current data for our sample, the flat or 
inverted spectral slopes between 888 MHz and 3 GHz implies that our sample might be similar in nature 
to that of \citet{Nyland2020}, i.e., dominated by young radio jets. 
On the other hand, \citet{Wolowska2021} suggested small changes of the accretion rate (by $\sim$30\%-40\%) may be sufficient to trigger low-power radio 
activity that evolves on the time-scale of decades. This may explain why most our sources have not shown large amplitude 
variability in the {optical and IR} light curves (Figure \ref{fig_optmir}). 
{Such a scenario can be tested with future works by identifying larger samples of galaxies with switched-on radio emission,  
especially for those with no or weak AGN activities. }
}

\begin{figure}[tb]
	\begin{minipage}[t]{0.48\linewidth}
		\centering
		\includegraphics[width=3.3in]{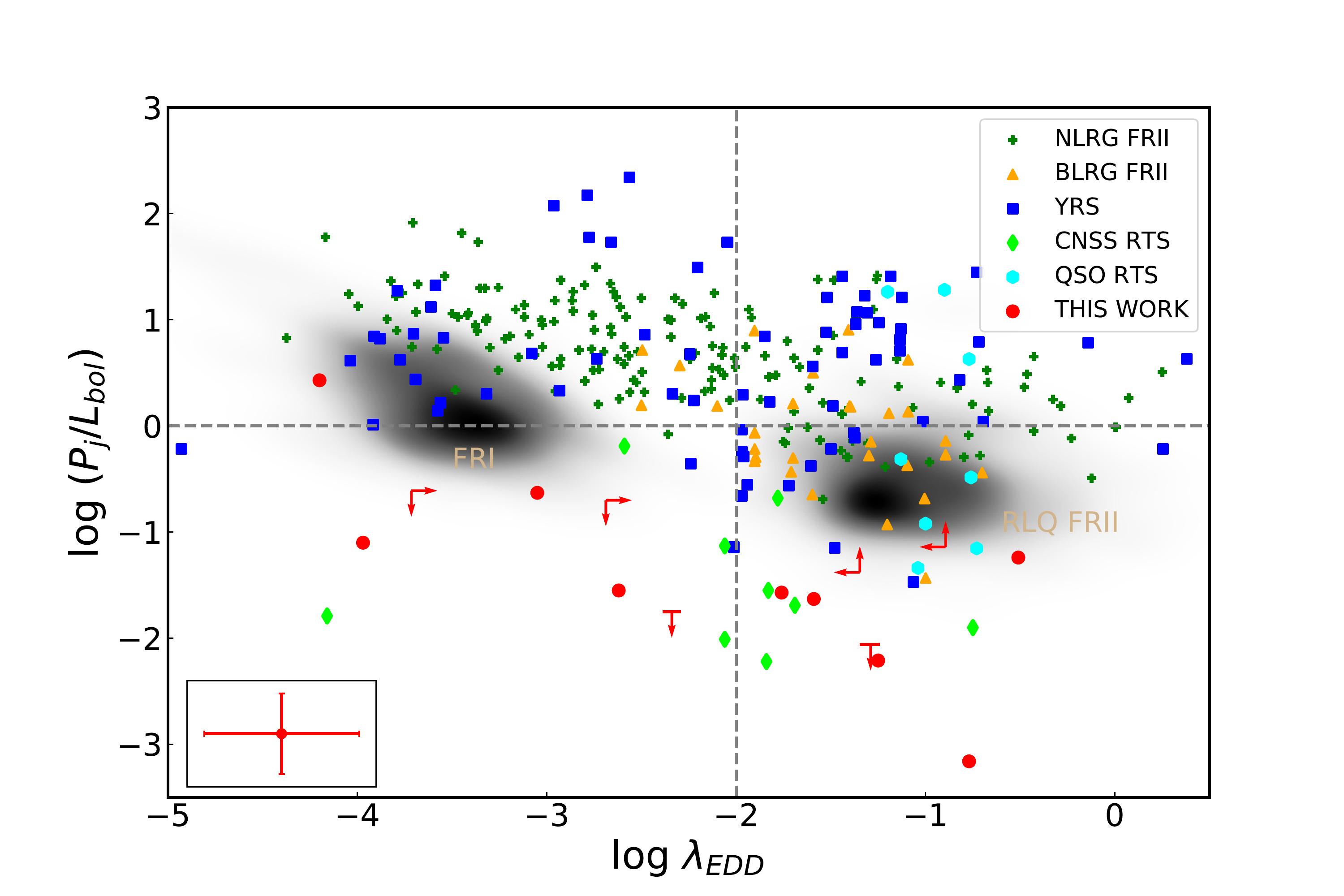}
	\end{minipage}
	\caption{
		Distributions of {the radio transients} in the $P_{j}/L_{\rm bol}-\lambda_{\rm Edd}$ plane, from our sample (red dots), 
	\citet{Nyland2020} (green dots) and \citet{Wolowska2021} (cyan dots). For comparison, we also show the distributions 
	from young radio sources (YRS) from \citet{Wojtowicz2020} and \citet{Liao2020}, and other types of AGNs analyzed by \citet{Rusinek2017},    including FR type II narrow-line radio galaxies (NLRG FRII), 
	broad-line radio galaxies (BLRG FRII), and radio-loud quasars (RLQ FRII, right density plot). 
	The FR type I radio galaxies from \citet{Capetti2017} are shown on the left density plot. 
	{The typical errors on $P_{j}/L_{\rm bol}$ and $\lambda_{\rm Edd}$ for our sample are shown in the left corner (Section 5.4 for the details). }
	}
	\label{Fig:jetpower}  
\end{figure}

In the context of new-born radio jets, it is interesting to compare the jet production efficiency of our sources with the sample of \citet{Wolowska2021} 
and \citet{Nyland2020}, as well as other galaxies with strong radio emission, which might be useful to understand 
the jet launching processes, and possibly the origin of the radio loudness of AGN activity. 
We first derived the jet kinetic power for our sample, $P_{\rm j}$, by using the {extrapolated 1.4 GHz flux densities (Figure \ref{Fig:fitslope})} and 
applying the scaling relation between the radio luminosity and {jet power \citep[e.g., Equation (4),][]{Wolowska2021}.\footnote{
Jet power $P_{\rm j}=\rm 5\times10^{22}\left(\frac{L_{\rm 1.4 GHz}}{W~Hz^{-1}}\right)^{6/7}$ erg s$^{-1}$, where $L_{\rm 1.4 GHz}$ is the rest-frame 1.4 GHz luminosity. }
}
Then the bolometric luminosity of the accretion disk, $L_{\rm bol}$, was calculated based on the {\sc [O iii]} luminosity with a  
correction factor of 3500 \citep{Heckman2004}.  
We considered the bolometric luminosity for the two star-forming galaxies in our sample as upper limit, as 
their {\sc [O iii]} emission is dominated by the star-forming activities. 
The jet production efficiency can be expressed as 
\begin{equation}
\eta_{\rm jet}=\frac{P_j}{\dot{M}_{\rm acc}c^2}=\frac{\eta_{\rm acc}P_j}{L_{\rm bol}}
\end{equation}
where $\dot{M}_{\rm acc}$ is the mass accretion rate,  and $\eta_{\rm acc}$ is the radiative efficiency 
of the accretion disk. We assumed the typical value of $\eta_{\rm acc}=$10\% in this paper when applicable. 
On the other hand, we parameterized the accretion rate using the Eddington ratio, $\lambda_{\rm Edd}=L_{\rm bol}/L_{\rm Edd}$, 
where $L_{\rm Edd}$ is the Eddington luminosity. 
It is suggested that an AGN could be transitioning between a radiatively efficient state and a radiatively inefficient state, 
if its accretion rate changes across $L_{\rm bol}/L_{\rm Edd}\sim10^{-2}$ \citep{Noda2018, Ruan2019}.

Figure \ref{Fig:jetpower} shows the distribution of our sources in the $P_{j}/L_{\rm bol}-\lambda_{\rm Edd}$ plane (red dots). 
Excluding {six} sources with either upper or lower limits, only one has the jet power in excess of the disk luminosity ($P_{j}/L_{\rm bol}>1$), 
and the remaining sources are distributed within a range 
from {log($P_{j}/L_{\rm bol})=-3.16$ up to $-0.63$}.  
{Furthermore}, the accretion rate in the sample spans a relatively wide range {between log($\lambda_{\rm Edd})=-4.2$ and $-0.5$}, 
with $\sim$50\% falling below $10^{-2}$. 
While the jet production efficiency and accretion rate {is} similar to the sample of \citet{Wolowska2021} (green dots), 
the distribution of the objects in the sample of \citet{Nyland2020} {appears to be} different (cyan dots). 
The latter sample occupies the locus with larger jet power and accretion rate, 
which overlaps with that of FR type II broad-line radio galaxies and radio quasars from the sample of \citet{Rusinek2017}. 
This is possibly due to the selection effect, as the sample of \citet{Nyland2020} includes only powerful quasars {with relatively high $\lambda_{\rm Edd}$}.  
In comparison to compact radio sources and GHz-peaked spectrum sources which are believed young radio galaxies, 
our sample tends to have a much lower jet production efficiency by at least an order of magnitude. 
For example, the {median $P_{j}/L_{\rm bol}$ is $\sim$0.1} for our sample, while 
it is 4.3 for the young radio galaxies. 
{On the other hand, there appears no dependence of jet production efficiency on the accretion rate. } 

{There are several {caveats} for the above comparison of jet power between different populations, thus the results 
should be treated with caution. 
First, the bolometric correction to {\sc [O iii]} luminosity is uncertain with a variance of 0.38 dex \citep{Heckman2004}, 
and the intrinsic scatter in the $M_{\rm BH}-\sigma_{\star}$ relation is 0.44 dex \citep{Gultekin2009}. 
These make difficult to robustly estimate the $L_{\rm bol}$ and $\lambda_{\rm Edd}$ in individual sources. 
As a result, we could only show the typical error bars of $\lambda_{\rm Edd}$ and $P_{j}/L_{\rm bol}$ in the left corner of Figure \ref{Fig:jetpower}. 
Second, the jet powers were calculated using the 1.4 GHz monochromatic radio luminosity. 
However, the scaling relation was originally derived using the radio luminosity at 151 MHz in FR II radio galaxies \citep[e.g.,][]{Willott1999}. 
It is not clear how the calculation of jet powers is affected by the changes of radio spectral shapes, 
{as the radio spectral index of $\alpha=0.8$ between 151 MHz and 1.4 GHz ($F_{\nu} \propto \nu^{-\alpha}$)
was assumed \citep{Rusinek2017}. In fact, jet power depends critically on the radio spectral index, {as well as} the 
upper and lower cut-off frequencies of the synchrotron spectrum \citep[Equation (1),][]{Willott1999}.} 
Third, since the {\sc [O iii]} emission comes from much extended regions ($\sim$kpc), it may not correspond to the same accretion power 
relevant to the brightened radio emission.  
{The quasi-simultaneous X-ray observations such as that obtained with Swift X-ray Telescope \citep{Burrows2005}}
might be useful to better constrain the accretion power related to the radio brightening, 
but such data are not available yet.  
Finally, although the {RACS} data at 888 MHz have been used to constrain the radio spectral shapes for our sources, 
it is still less accurate and insufficient to localize the peak frequencies. 
Quasi-simultaneous multi-frequency radio observations are highly encouraged for confirming the peaked radio spectrum, 
one of critical characteristics of young radio jets, which will help to further understand the nature of 
radio transients revealed in VLASS. 
}

\section{Conclusion} \label{sec:Conclusion}

{By analyzing the galaxies observed by FIRST and VLASS, we have identified a sample of 18 slow-evolving radio transients at $z\simlt0.3$, 
characterized by radio variability (brightening in VLASS epoch I observations) on decadal timescales. 
The location of the radio emission is consistent with the optical center of the galaxy from SDSS photometry, 
with positional offsets less than 0$\farcs$41, indicating an origin from the nucleus, perhaps associated with the 
accreting SMBHs. 
All these galaxies are detected in the second epoch VLASS observations, and the 3 GHz radio emission 
does not show significant variations over $\sim$30-36 months ($\simlt50$\%).  
By checking for the archival {RACS} data, we find 15 out of 18 sources are observed and an inverted radio spectrum between 
888 MHz and 3 GHz can be inferred, suggesting an origin from optically thick regions. 
Based on the SDSS spectroscopy data, the 18 galaxies can be classified as LINERs (8), star-forming galaxies (2), 
passive galaxies (2), composites (2), and AGNs (4). 
Most of the host galaxies are intrinsically red and massive, consistent with them being early-type galaxies from morphological analysis.  
Except for one source, we do not detect any optical and mid-IR flares in CRTS, ZTF, and WISE light curves spanning $\sim$8--12 yr 
before the start of the VLASS survey. 
The combination of radio luminosities, variability amplitude and timescales, spectral shapes, and host galaxy properties 
rules out the origins from stellar explosions such as SNe and GRBs, as well as intrinsic variability of radio-quiet AGN. 
Non-detections of optical and mid-IR flares, as well as their relatively large black hole masses ($M_{\rm BH}\simgt10^{8}$\msun) 
suggest that most sources in our sample are not TDEs. 
Young radio jet remains a likely scenario, which can be tested with further multi-frequency 
{as well as high spatial resolution radio observations of galaxies in the sample}. 
 


}

 
\acknowledgments{
{We thank the anonymous referee for helpful comments that improved this work. }
The National Radio Astronomy Observatory is a facility of the
National Science Foundation operated under cooperative agreement
by Associated Universities, Inc.
This work makes use of data products from the Wide-field Infrared Survey Explorer, which is a joint project of 
the University of California, Los Angeles, and the Jet Propulsion Laboratory/California Institute of Technology, funded by the National Aeronautics and Space Administration. 
This work makes use of data products from Catalina Real-time Transient Survey, and 
the Zwicky Transient Facility Project, which is supported by the National Science Foundation under Grant No. AST-1440341. 
{The Australian SKA Pathfinder is part of the Australia Telescope National Facility which is managed by CSIRO. Operation of ASKAP is funded by the Australian Government with support from the National Collaborative Research Infrastructure Strategy. ASKAP uses the resources of the Pawsey Supercomputing Centre. Establishment of ASKAP, the Murchison Radio-astronomy Observatory and the Pawsey Supercomputing Centre are initiatives of the Australian Government, with support from the Government of Western Australia and the Science and Industry Endowment Fund. We acknowledge the Wajarri Yamatji people as the traditional owners of the Observatory site.
This paper includes archived data obtained through the CSIRO ASKAP Science Data Archive, CASDA (http://data.csiro.au).
Funding for the Sloan Digital Sky 
Survey IV has been provided by the 
Alfred P. Sloan Foundation, the U.S. 
Department of Energy Office of 
Science, and the Participating 
Institutions. 
SDSS-IV acknowledges support and 
resources from the Center for High 
Performance Computing  at the 
University of Utah. The SDSS 
website is www.sdss.org.
}
 The work is supported by Chinese NSF through grant No. 11822301, 12192220, 12192221, and 11833007. 
}

\software{CASA (v5.3.0; McMullin et al. 2007), CIGALE (Boquien et al. 2019)}

\bibliographystyle{mnras}
\bibliography{ref}

\appendix

\section{optical spectrum and image cutouts for individual sources} 

\setcounter{figure}{0}
\renewcommand{\thefigure}{A\arabic{figure}}

{In Figure A1}, we present the SDSS optical spectrum of each galaxy in our sample, 
the SDSS r-band optical image, FIRST image,  VLASS epoch I image and VLASS epoch II image.
\begin{figure*}[htbp]
\centering
\includegraphics[scale=0.3]{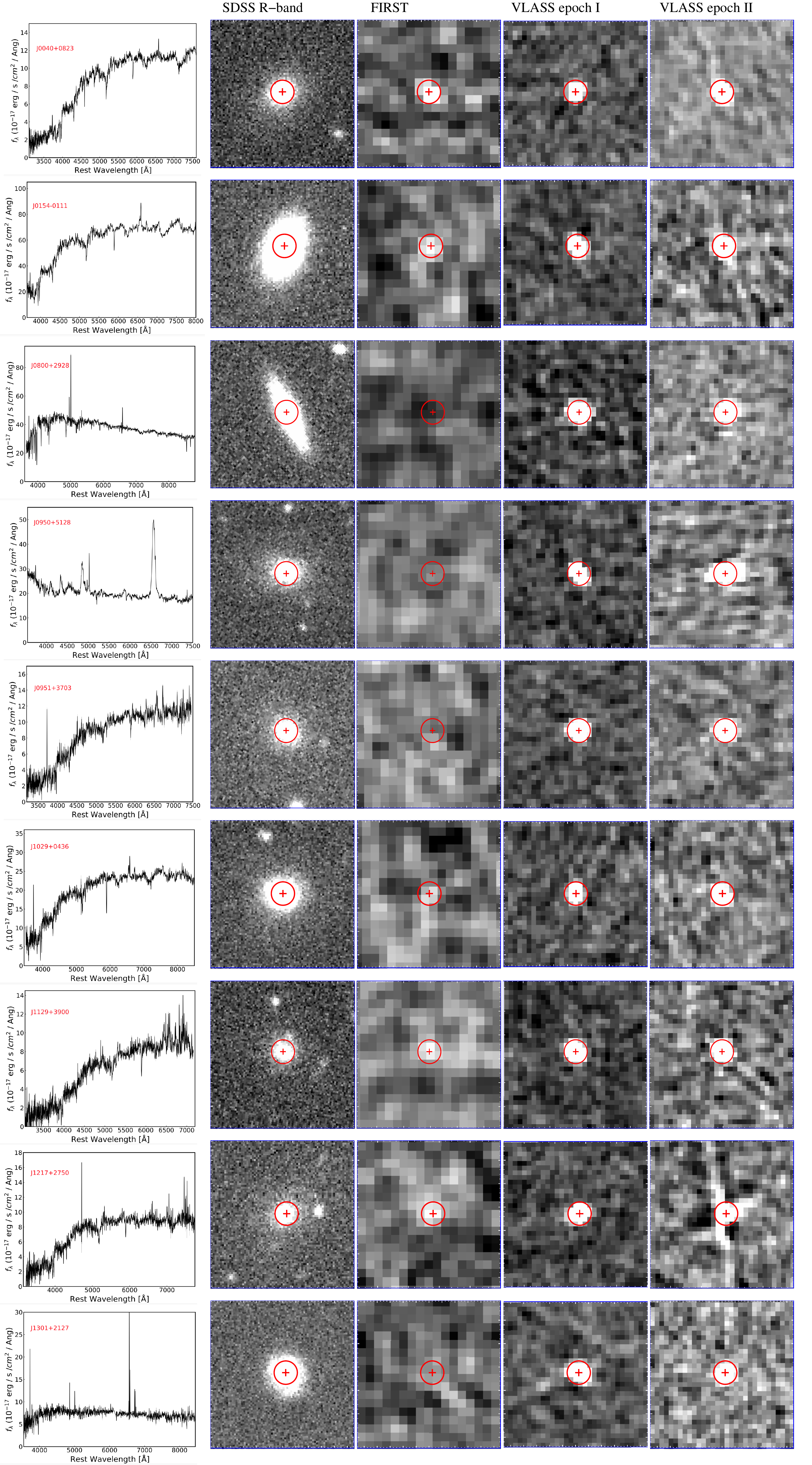}
\caption{
From left to right, SDSS optical spectrum of host galaxy, SDSS r-band image, 
FIRST image, VLASS epoch I image and VLASS epoch II image.  
The red circle has a radius of $r=2\farcs$5,  comparable to the beam size of VLASS images.  
The red cross represents the center of radio emission from VLASS epoch I observations. 
Each image cutout has a size of $30\arcsec\times30\arcsec$.  
 }
\end{figure*}

\setcounter{figure}{0}

\begin{figure*}[htbp]
\centering
\includegraphics[scale=0.3]{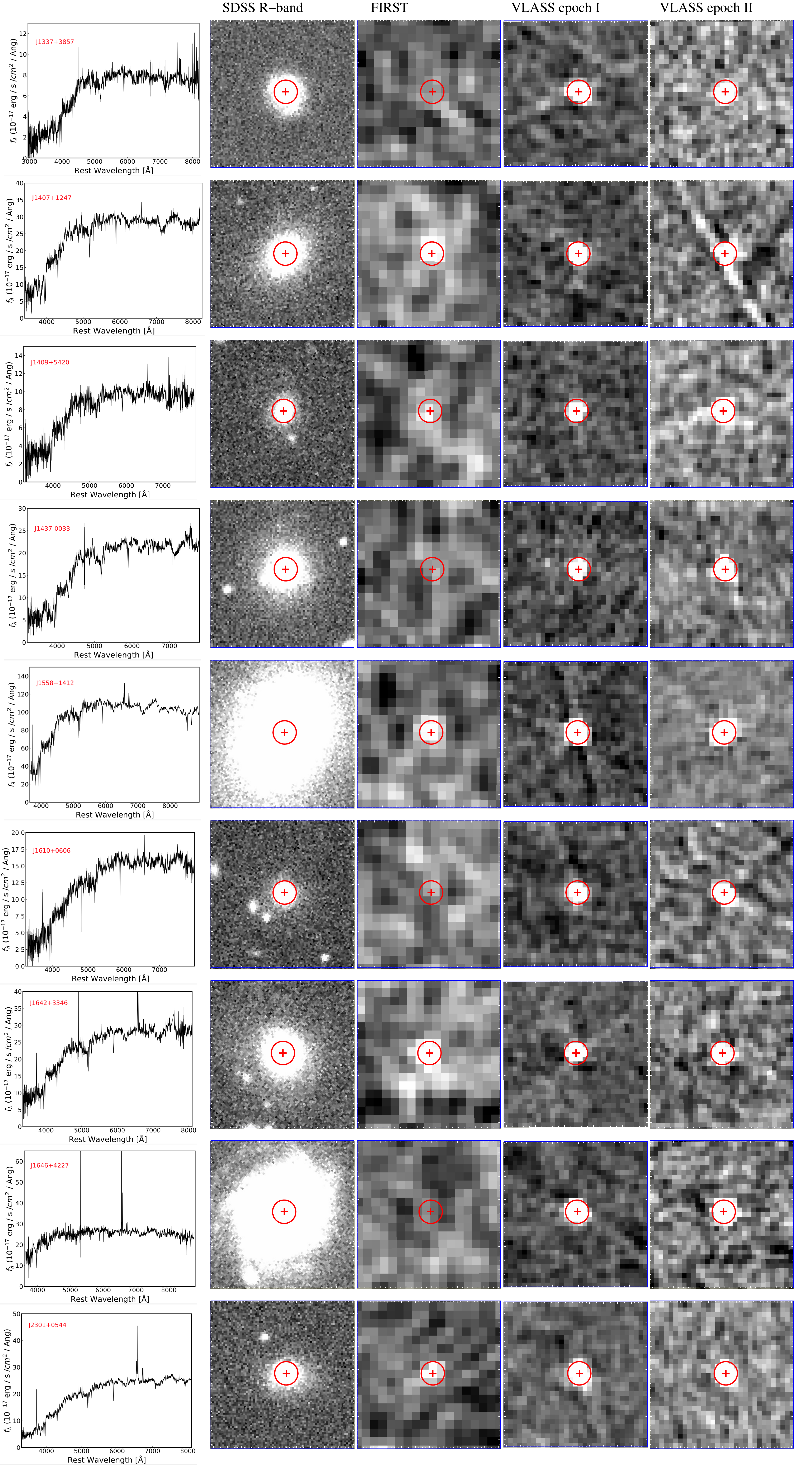}
\caption{
 {Continued.}
 }
\end{figure*}

\section{{Comparison of radio flux variations between different samples}} 
\setcounter{figure}{0}
\renewcommand{\thefigure}{B\arabic{figure}}

{
Following the approaches presented in Section 3 (Figure 1), we retrieved the VLASS 
epoch I and II observations and measured the 3 GHz fluxes to 
compute the variability amplitudes for the sample of \citet{Nyland2020} and \citet{Wolowska2021}.  
The former sample includes only quasars that may have transitioned from radio-quiet to radio-loud in VLASS, 
while the latter sample contains the galaxies with transient radio emission selected in the CNSS data. 
The constant or small flux variations at 3 GHz of our sample over timescales of 30-35 months are 
consistent with that presented in \citet{Nyland2020} and \citet{Wolowska2021}, as shown in {Figure B1}.  
It should be noted that the two works either did not use the VLASS data or used only epoch I data. 
Our uniform analysis of the data from VLASS epoch I and II observations makes the above comparison fair. 

}

\begin{figure*}[bpt]
	\begin{minipage}[t]{0.48\linewidth}
		\centering
		\includegraphics[width=3.5in]{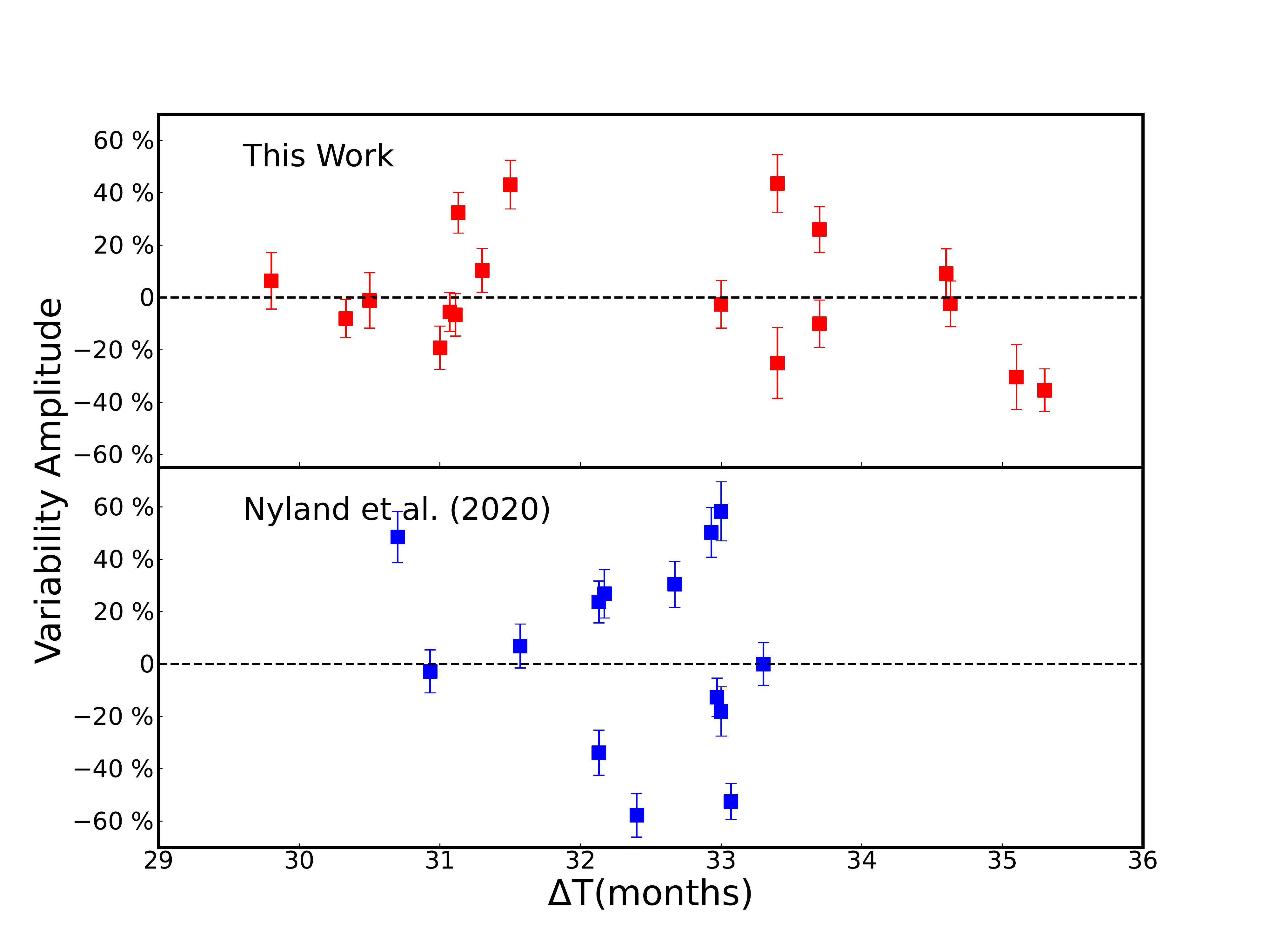}
	\end{minipage}
	\begin{minipage}[t]{0.48\linewidth}
		\centering
		\includegraphics[width=3.5in]{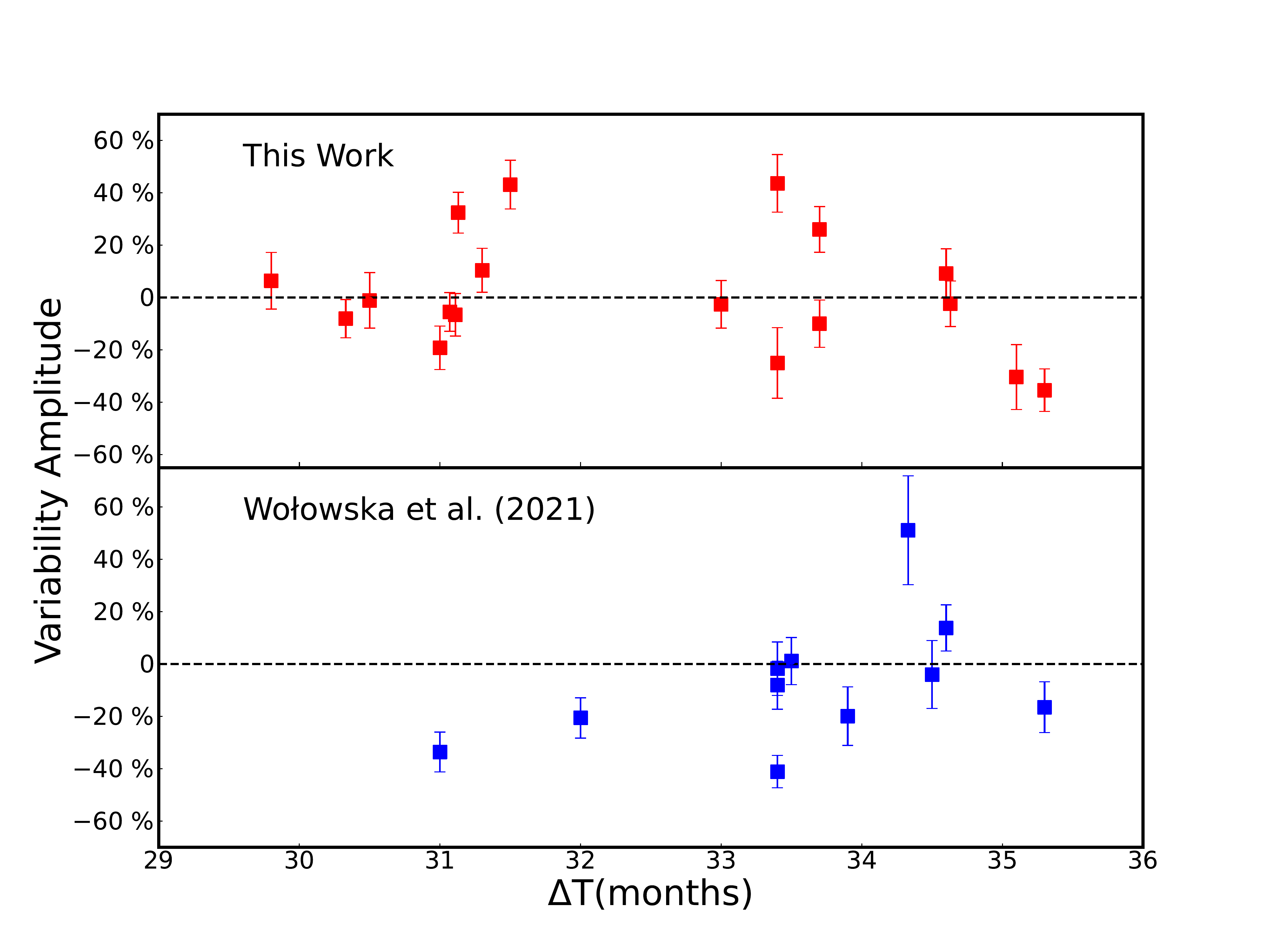}
	\end{minipage}
	\caption{{\it Left panel:} The variability amplitude of radio flux between VLASS epoch I and epoch II observations
	 for our sample (upper panel) and the sample presented by \citet{Nyland2020} (lower panel). 
	{\it Right panel:}  The same as left, but for the comparison with the sample presented by \citet{Wolowska2021}. }
	 \vspace{0.5cm}
\end{figure*}

\section{{RACS} images at 888 MHz} 
\setcounter{figure}{0}
\renewcommand{\thefigure}{C\arabic{figure}}

{ 
{Figure C1} shows the radio images at 888 MHz for 15 out 18 sources observed in the 
Rapid ASKAP Continuum Survey \citep[RACS,][]{McConnell2020}. 
The remaining three sources (J0950, J1409, and J1646) are 
not covered by the {RACS} survey due to their too high declination (DEC$>$+40). }


\begin{figure*}[htbp]
\centering
\includegraphics[scale=0.68]{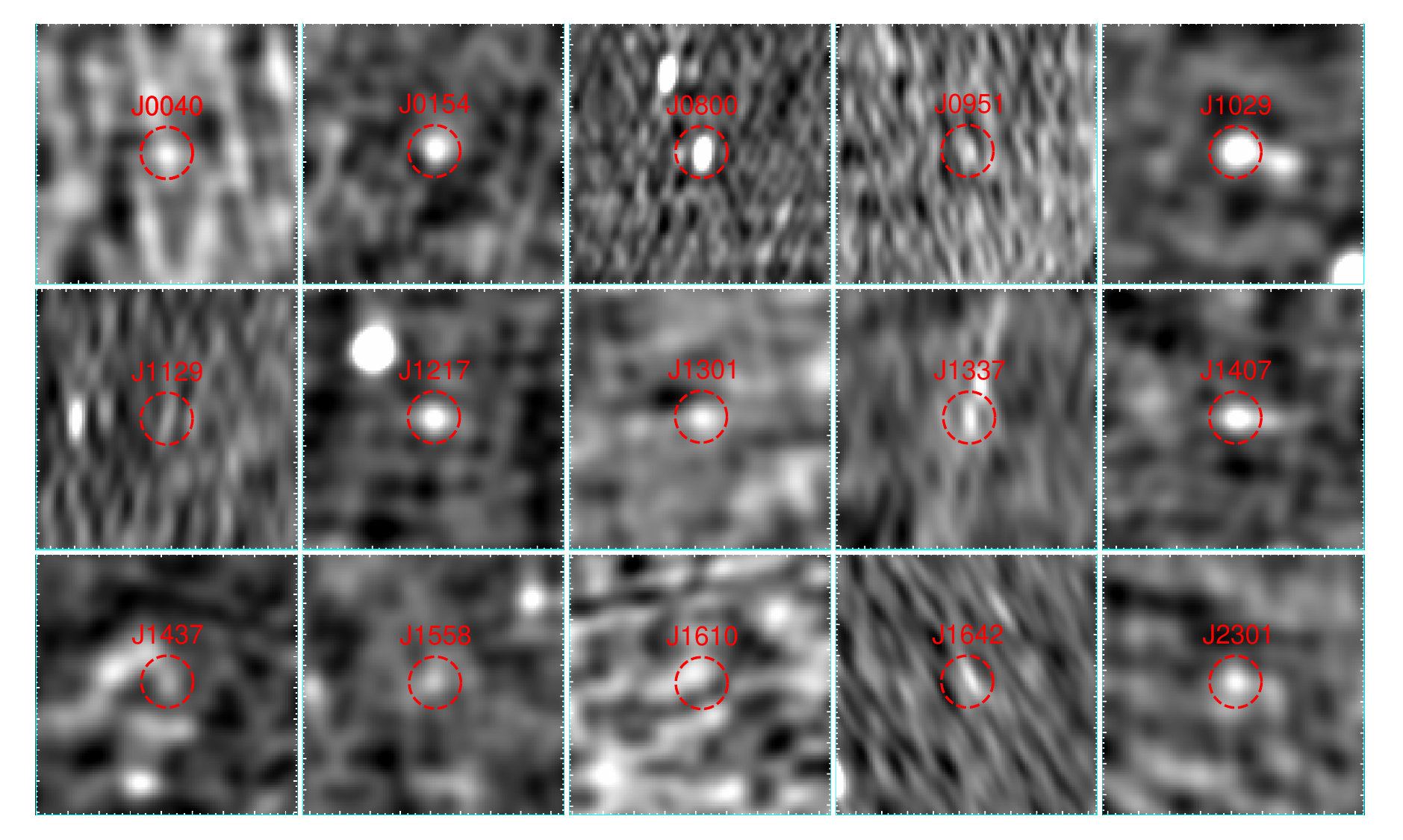}
\caption{
The radio images at 888 MHz for 15 sources observed in the Rapid ASKAP Continuum Survey.  
The red circle in each panel has a radius of 25\arcsec. The image has a size of 4\arcmin$\times$4\arcmin. 
 }
\end{figure*}

\section{optical spectral fittings and SED fittings to the SDSS and WISE photometry}  
\setcounter{figure}{0}
\renewcommand{\thefigure}{D\arabic{figure}}

Three objects in our sample, including J0040+0823, J1337+3857 and J2301+0544, are not listed in either the MPA-JHU catalog 
or combined SDSS and WISE photometric catalog \citep{Chang2015}, because their SDSS spectra were observed later than 2010.
Thus, we measured the properties of their host galaxies by modeling their SDSS spectra and SEDs, as shown in {Figure D1}. 
We modeled the SDSS spectra following \citet{Dong2012}.
In brief, we modeled the continuum by fitting the spectra in regions unaffected by emission lines with a linear combination of simple stellar population templates from \citet{Bruzual2003}.
The templates were broadened with a Gaussian function, of which the $\sigma$ value was used as the measurement of stellar velocity dispersion.
The results are shown in the left column of {Figure D1}.
After subtracting the continuum model, we then fitted the H$\beta$, [OIII] $\lambda\lambda$4959, 5007, H$\alpha$ and [NII] $\lambda\lambda$6548, 6583 emission lines with the following assumptions: (1) all the emission lines can be represented with single Gaussians with redshifts and width (in velocity) tied together; 
(2) the flux ratios of {\sc [O iii]} and {\sc [N  ii]} doublets are fixed at 3.
The results are shown in the middle two columns of {Figure D1}.
To obtain the stellar mass, we constructed the SEDs of the galaxies using the photometric data from the SDSS and WISE surveys, and fitted the SED using the CIGALE code \citep{Boquien2019}.
In the fittings we also assumed that the star-formation history is an exponential function and the Salpter's initial mass function was adopted.
The results are shown in the right column of {Figure D1}.

\begin{figure*}[htbp]
\centering
\includegraphics[scale=0.8]{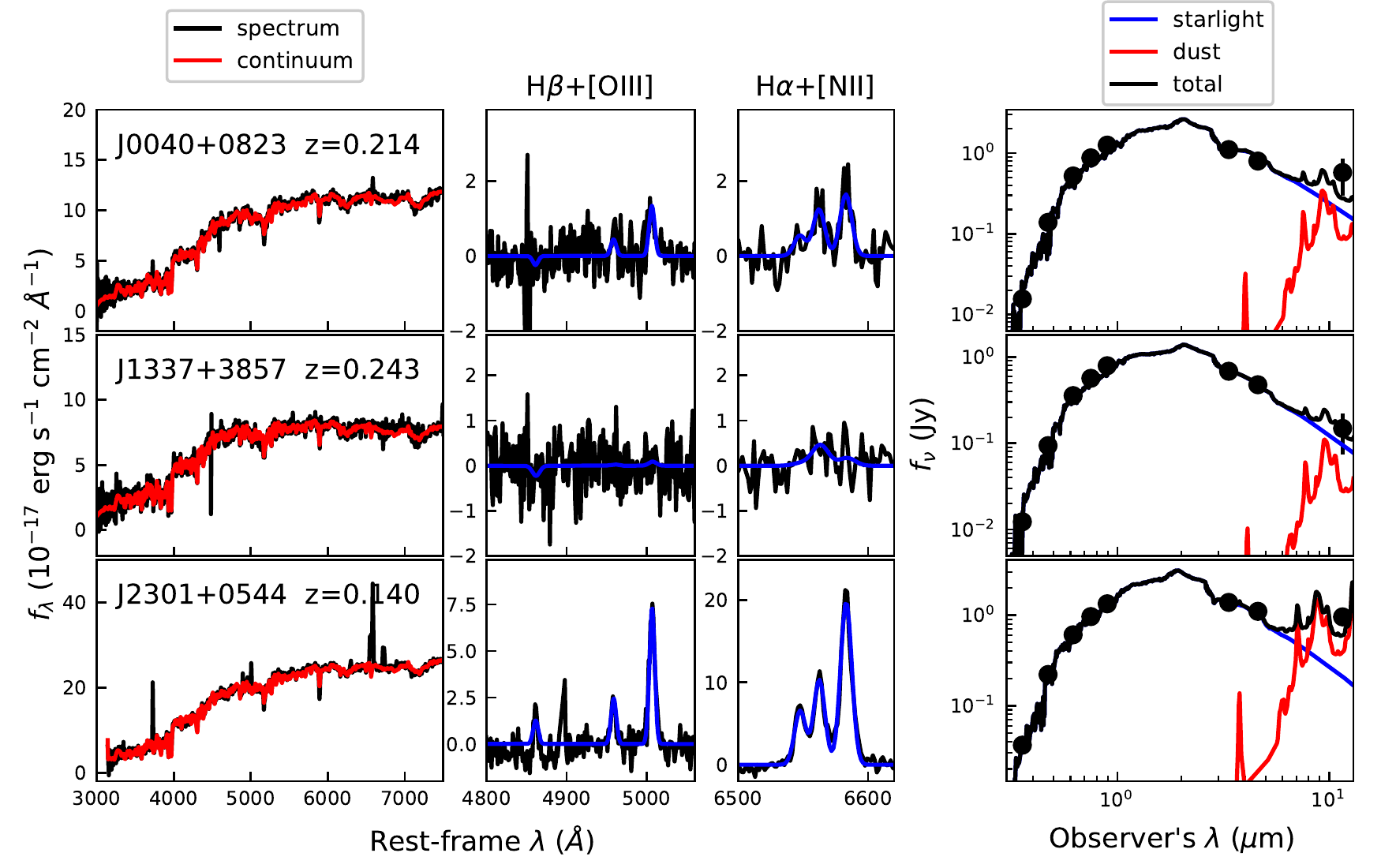}
\caption{
Illustration of the continuum and emission-line fittings of the SDSS spectrum (left panel). 
In the middle two panels, we show a zoomed-in view of the emission-line profile fittings for 
the H$\beta$+{\sc [O iii]} region, and H$\alpha$+{\sc [N ii]} region, respectively. 
The SED fitting results from CIGALE are shown on the right panel. 
 }
\end{figure*}

\end{document}